\documentclass[journal,12pt,onecolumn,draftclsnofoot]{IEEEtran}
\usepackage{amsmath,amssymb,mathtools}
\usepackage{array}
\usepackage{cite}
\usepackage{url}
\newcommand{\cnt}{\mathrm{cnt}}
\newcommand{\loc}{\mathrm{loc}}
\newcommand{\OPP}{\mathrm{OPP}}
\newcommand{\ML}{\mathrm{ML}}
\newcommand{\degns}{\deg_{\mathrm{NS}}}
\newcommand{\Q}{\mathbb{Q}}
\newcommand{\F}{\mathbb{F}}
\newcommand{\Zn}{\mathbb{Z}_n}
\newcommand{\ZM}{\mathbb{Z}_M}
\newcommand{\Bool}[1]{B_{#1}}

\newcommand{\Emix}[2]{E^{\mathrm{#1}}_{#2}}
\newcommand{\dmin}{d_{\min}}
\newcommand{\cw}{\mathcal{C}}
\newcommand{\Tgt}{\mathcal{T}}

\newtheorem{theorem}{Theorem}
\newtheorem{proposition}[theorem]{Proposition}
\newtheorem{lemma}[theorem]{Lemma}
\newtheorem{corollary}[theorem]{Corollary}
\newtheorem{conjecture}{Conjecture}
\newtheorem{problem}{Open Problem}
\newtheorem{example}{Example}
\newcounter{remark}
\newenvironment{remark}[1][]{\refstepcounter{remark}\par\smallskip\noindent
  \textit{Remark~\theremark\if\relax\detokenize{#1}\relax\else\ (#1)\fi:}\ }{\par\smallskip}

\begin{document}

\title{Encoding-Sensitive Nullstellensatz Certificates\\ for Syndrome Exclusion in LDPC Codes:\\ Sharp Local Bounds and a Linear Separation}

\author{Malihe~Aliasgari,
        Hassan~Khodaiemehr,
        and~Khadijeh~Bagheri%
\thanks{M.~Aliasgari is with the School of Integrative Science and Technology, Department of Mathematical Sciences, Kean University, Union, NJ, USA.}%
\thanks{H.~Khodaiemehr and K.~Bagheri are with the School of Engineering, University of British Columbia, Okanagan Campus, Kelowna, BC, Canada.}%
\thanks{A preliminary version of this work~\cite{itw27} was submitted to the 2027 IEEE Information Theory Workshop (ITW).}%
}

\markboth{}{}

\maketitle

\begin{abstract}
Certifying that a parity-check matrix has no nonempty set of at most $t$ variable nodes leaving at most $b$ checks unsatisfied arises in minimum-distance and trapping-set analysis. Such absence is the infeasibility of a polynomial system, provable by an independently checkable Nullstellensatz certificate. The degree of these certificates depends strongly on how the two cardinality constraints are encoded, although the encodings have identical solutions. For a local encoding with inclusion--exclusion and cardinality obstructions, we construct explicit certificates of degree at most $t+b+K+1$, with $K$ logarithmic in the maximum check degree; for fixed $t$, $b$, and bounded check degree, this bound is independent of block length and is tight in general. For a standard counting encoding, a generalized planted-knapsack theorem, which pulls back a knapsack functional with an arbitrary target, gives lower bounds linear in the block length for cycle codes, regular codes of girth at least six near the feasibility boundary, array-based quasi-cyclic codes and minimum-distance certificates for arrays of circulant permutation matrices, where the local degree stays bounded. On array codes and regular codes of girth at least eight, planted column pairs extend these bounds to odd column weight with no unsatisfied checks. The planting argument is optimal among such pullbacks. All computed degrees and certificates are verified in exact rational arithmetic; bounds are distinguished from exact minimum degrees. Local certificates for quasi-cyclic matrices, including a lifted fifth-generation New Radio base graph, pass two independent verifiers.
\end{abstract}


\section{Introduction}
\IEEEPARstart{S}{mall} variable-node supports that leave few parity checks unsatisfied govern much of the finite-length behaviour of low-density parity-check (LDPC) codes. For a binary parity-check matrix $H$ and a set $S$ of variable nodes, let $\Phi(S)$ denote the number of parity checks with odd induced degree on $S$. Supports with $\Phi(S)=0$ are codewords, and small supports with small nonzero $\Phi(S)$ are closely related to the structural trapping sets behind error floors of iterative decoders~\cite{richardson03,dolecek10}. Related objects include stopping sets~\cite{di02}, absorbing sets~\cite{dolecek10} and graph-cover pseudocodewords~\cite{vontobel05}, and efficient exhaustive procedures locate small stopping sets and elementary trapping sets in a given Tanner graph~\cite{wang09,karimi12,raeisi24,rosnes09,hashemi15}.

We consider the following bounded-support syndrome-exclusion problem. Given $H$, a support-size bound $t$ and an unsatisfied-check bound $b$, decide whether there is a support $S$ with
\begin{equation}
1\le |S|\le t,\qquad \Phi(S)\le b .\label{eq:problem}
\end{equation}
This is the union of the structural $(a,\beta)$ trapping-set classes with $a\le t$ and $\beta\le b$. For $b=0$ it asks for a nonzero codeword of weight at most $t$, so infeasibility of~\eqref{eq:problem} with $b=0$ certifies $\dmin\ge t+1$. We do not propose a new trapping-set class or a faster search; we ask how the \emph{polynomial formulation} of the same exclusion problem affects the complexity of an algebraic proof that no such support exists.

If a polynomial system $E\subset\Q[\mathbf v]$, including Boolean axioms, has no common zero over $\overline{\Q}$, Hilbert's Nullstellensatz gives polynomials $a_g\in\Q[\mathbf v]$ with
\begin{equation}
1=\sum_{g\in E}a_g\,g .\label{eq:ns}
\end{equation}
Such an identity is a certificate of infeasibility: once generated, it can be checked by exact polynomial arithmetic, without trusting the program that produced it. Its complexity is measured by $\max_g\deg(a_gg)$, the standard degree measure of algebraic proof complexity~\cite{beame96,pitassi97,miksa24}. Proof degree can depend strongly on the formulation even when two systems have the same Boolean solutions: degree lower bounds are known for parity-type systems~\cite{grigoriev98} and for subset-sum (knapsack) systems~\cite{ips99,grigoriev01}, and redundant constraints can change proof degree substantially~\cite{deloera11}.

We compare two exact encodings of~\eqref{eq:problem} with identical Boolean solutions. The \emph{counting encoding} introduces counters and one-hot selectors for the support size and the syndrome weight; its number of generators grows linearly with the dimensions of $H$. The \emph{local encoding} uses a truncated inclusion--exclusion polynomial and explicit cardinality obstructions that do not depend on $H$; all dependence on $H$ stays in the parity generators. The instances on which we prove large degree are often easy combinatorially: at $t=1$ and $b\le w-1$, where $w$ is the column weight, infeasibility follows from the column weight alone, and the minimum distances of our array-code examples are found by a short exhaustive search. The point is that the formulation decides whether a low-degree algebraic proof exists at all: on the same infeasible instances the counting encoding needs degree linear in the block length, while the local encoding has a certificate of bounded degree for fixed $t$, $b$ and bounded check degree. This is a proof-complexity phenomenon with consequences for algebraic certification; it does not give a decoding or search algorithm.

\subsection{The Conference Version}
A preliminary version of this work~\cite{itw27} (i)~introduced the two encodings and proved them exact; (ii)~constructed explicit local certificates of degree at most $t+b+K+2$, where $K$ depends logarithmically on the maximum check degree; (iii)~established a \emph{finite} separation on cycle codes $C_n$: for $n=10,12,14,16$ the counting encoding needs degree at least $5,5,6,7$, whereas the local construction gives degree at most~$4$; (iv)~reduced the counting-encoding degree on $C_n$, up to a factor of two, to a compact Boolean system $\OPP_n$ and conjectured linear growth of its degree; and (v)~generated local certificates for regular and irregular QC-LDPC matrices and verified them with two independent exact-arithmetic implementations. Asymptotic analysis, lower bounds beyond cycle codes and minimum-distance certification were outside the scope of~\cite{itw27}. The remaining results are new to this paper; Section~\ref{sec:position} explains which parts rest on standard techniques.

\subsection{Contributions}
\begin{enumerate}
\item \emph{Sharp constructive local certificates} (Section~\ref{sec:local}). The local encoding admits an explicitly constructed certificate of degree at most $t+b+K+1$, one less than in~\cite{itw27}, with $O(n^t)$ terms (Theorem~\ref{thm:A}). For fixed $t$, $b$ and maximum check degree the bound is uniform in the block length, and it cannot be improved in general (Proposition~\ref{prop:tight}). At $(t,b)=(1,0)$ the local degree is exactly two on column-regular matrices whose checks have degree two or three (Proposition~\ref{prop:B}).
\item \emph{A generalized planted-knapsack theorem} (Section~\ref{sec:planting}). If a few columns are fixed to one and $M$ disjoint sets of columns carry new Boolean variables, a substitution with constant selectors maps the counting encoding onto a knapsack with an arbitrary rational target. Pulling back the corresponding functional (Lemma~\ref{lem:A}) gives $\degns\ge M+1$ under four conditions on $H$ and the planted sets, one of them a linear equation (Theorem~\ref{thm:B}). The general target removes the restriction $|O|\le t+1$ of single-column plantings, and planted sets of even size remove the parity obstruction for odd column weight (Lemma~\ref{lem:G}).
\item \emph{Linear separations} (Section~\ref{sec:sep}). Theorem~\ref{thm:B} gives counting-degree lower bounds linear in the block length for cycle codes, for regular families of girth at least six near the feasibility boundary, for quasi-cyclic (QC) and array codes, and through planted $6$-cycles for column weight three at $t\in\{2,3\}$. With a generic $O(n)$ upper bound these are $\Theta(n)$-versus-$O(1)$ separations for fixed $t$ and $b$ and bounded check degree (Table~\ref{tab:compare}).
\item \emph{Minimum-distance certification} (Section~\ref{sec:md}). If $O$ is a codeword with $s$ check-disjoint columns removed and the planted sets are single columns, the counting degree for certifying $\dmin\ge t+1$ is at least $M+1$ for every $t$ with $\dmin-2s\le t<\dmin$ (Theorem~\ref{thm:H}). For $w\times L$ arrays of circulant permutation matrices with girth at least six and $L\ge w+1$ this is a $\Theta(n)$-versus-$O(1)$ separation for certifying the minimum distance (Corollary~\ref{cor:md-asym}). On $H(3,5,47)$ the counting degree is at least $45$ for every $t$ from $2$ to $5$, against local degree at most $8$, from a perfect packing of $44$ planted columns (Example~\ref{ex:A}).
\item \emph{Odd column weight} (Section~\ref{sec:odd}). Planted pairs of columns attached to an induced path or a star give linear lower bounds at $(t,b)=(1,b)$, $b\le w-1$, for every column weight (Corollary~\ref{cor:E}). We state the Tanner-graph conditions precisely, show that girth six alone does not suffice, and settle the odd-weight question for the array codes $H(w,L,Z)$ and for regular families of girth at least eight (Corollary~\ref{cor:odd}).
\item \emph{Mixed encodings, optimality and limits} (Sections~\ref{sec:mixed} and~\ref{sec:limits}). On cycle codes at $(t,b)=(1,0)$ the growth requires both cardinalities to be counted (Proposition~\ref{prop:D}). The knapsack target has degree exactly $M+1$, no pullback of this kind proves more (Theorem~\ref{thm:cap}, Corollary~\ref{cor:cap-planted}), and the bound $M+1$ is attained for every $M$ (Proposition~\ref{prop:sharp}), for $M\ge3$ also when every planted check contains a second column (Proposition~\ref{prop:partner-cex}).
\item \emph{Exact verification} (Section~\ref{sec:verif}). All computational degree claims and certificate identities are verified in exact rational arithmetic, and lower bounds are kept distinct from exact minimum degrees; packing optima are proven with CP-SAT and confirmed with a second solver, and the QC certificates are accepted by two independent verifiers. Scripts, configurations, the superseded square-root bound for cycle codes and the proof of Proposition~\ref{prop:partner-cex} are in the supplementary material.
\end{enumerate}

\subsection{Positioning and Scope}\label{sec:position}
The lower bounds use a restriction argument, a standard way of transferring degree lower bounds in algebraic proof complexity: a degree-non-increasing substitution carries a dual object, a design in the sense of~\cite{beame96,buss98}, for a knapsack system to the counting encoding. Knapsack lower bounds themselves are classical~\cite{ips99,grigoriev01}, and the generic upper bound of Proposition~\ref{prop:generic-upper} is folklore~\cite{pitassi97}. What is specific to LDPC codes is where the plantings come from: codewords, induced paths, stars, planted pairs and $6$-cycles in Tanner graphs, the conditions on $H$ under which they exist, and the analysis of what this method can and cannot prove.

The local encoding pays for low degree with $\binom{n}{t+1}+\binom{m}{b+1}$ extra generators, and its certificate has $O(n^t)$ terms, the same order as exhaustive enumeration of the supports of size at most $t$. The construction is therefore not presented as faster than enumeration: search algorithms~\cite{wang09,karimi12,raeisi24,rosnes09,hashemi15}, integer programming or CP-SAT are more appropriate when the goal is only to decide whether a harmful support exists. What a certificate adds is a proof object that a generic polynomial-identity checker can verify. Lower proof degree does not by itself mean lower computational cost, and we draw no conclusions about Gr\"obner-basis running times. All statements are over $\Q$, and several depend on characteristic zero (Remark~\ref{rem:char}).

\subsection{Related Work}
Exhaustive or structured search decides whether stopping sets, trapping sets, absorbing sets or low-weight codewords exist~\cite{dolecek10,wang09,karimi12,raeisi24,rosnes09,hashemi15}, and graph covers give another framework for finite-length iterative decoding~\cite{vontobel05}. Array codes as LDPC codes were introduced by Fan~\cite{fan00}, and their minimum distance has been studied in~\cite{yang03,rosnes14}; arrays of circulant permutation matrices have minimum distance at most $(w+1)!$ when they have at least $w+1$ column blocks~\cite{mackay01,fossorier04}, a bound that Smarandache and Vontobel derive from a permanent bound for quasi-cyclic codes~\cite[Th.~8, Cor.~9]{sv12}. Finite-geometry LDPC codes~\cite{kou01} provide girth-six Tanner graphs in which every two columns share a check. Gr\"obner bases and binomial ideals associated with linear codes have been used for minimum-support codewords, test sets and algebraic decoding~\cite{borges08,marquez16,bulygin09}; we study instead the degree of algebraic \emph{proofs of nonexistence} for a fixed exclusion problem. Encoding sensitivity of Nullstellensatz and polynomial-calculus degree is well established in proof complexity~\cite{deloera11,lauria17,derezende21}, and independently checkable proof objects are standard in SAT, pseudo-Boolean reasoning, integer programming and polynomial-calculus verification~\cite{gocht21,cheung17,kaufmann24}.

\section{Problem Formulation and Exact Encodings}\label{sec:prelim}

\subsection{Syndrome Exclusion and Parity}
Let $H\in\F_2^{m\times n}$ have row supports $N(j)=\{i:H_{ji}=1\}$ and check degrees $d_j=|N(j)|$, $j\in[m]$, and column supports $C(i)=\{j:H_{ji}=1\}$, whose size $|C(i)|$ is the column weight. For $A\subseteq[n]$ write $C(A)=\bigcup_{i\in A}C(i)$ and $c_j(A)=|N(j)\cap A|$, so that $\Phi(A)=\#\{j:c_j(A)\text{ odd}\}$. For $x\in\{0,1\}^n$ with support $S(x)=\{i:x_i=1\}$, check~$j$ has induced degree $s_j(x)=\sum_{i\in N(j)}x_i$ and odd-check indicator $u_j=s_j(x)\bmod 2$, so that $\Phi(S(x))=\sum_ju_j$. For integers $1\le t\le n$ and $b\ge0$, problem~\eqref{eq:problem} reads
\begin{equation}
1\le \textstyle\sum_{i=1}^{n}x_i\le t,\qquad \sum_{j=1}^{m}u_j\le b ,\label{eq:feas}
\end{equation}
and we certify its infeasibility. Parity must be expressed polynomially over~$\Q$, and $s_j-u_j=2q_j$ with a free rational $q_j$ is insufficient: $s_j=2$ admits $(u_j,q_j)=(1,\tfrac12)$. We therefore use Boolean quotient bits $q_{j,0},\dots,q_{j,K_j-1}$ with
\begin{equation}
K_j=\big\lceil\log_2(\lfloor d_j/2\rfloor+1)\big\rceil,\qquad Q_j=\textstyle\sum_{k=0}^{K_j-1}2^kq_{j,k},\label{eq:Kj}
\end{equation}
the parity generators
\begin{equation}
L_j=\sum_{i\in N(j)}x_i-u_j-2Q_j,\qquad j\in[m],\label{eq:parity}
\end{equation}
and the Boolean axioms
\begin{equation}
\Bool{v}=v^2-v\quad\text{for every Boolean variable } v.\label{eq:bool}
\end{equation}
At a Boolean zero of~\eqref{eq:parity}, $s_j-u_j=2Q_j$ is even and nonnegative, so $u_j=s_j\bmod 2$ and $Q_j=\lfloor s_j/2\rfloor$, whose bits exist and are unique because $2^{K_j}-1\ge\lfloor d_j/2\rfloor$; for $d_j\le1$, $K_j=0$ and $L_j=s_j-u_j$. We write $K=\max_jK_j$, $d_{\max}=\max_jd_j$ and $N_t(n)=\sum_{s=0}^{t}\binom ns$.

\subsection{The Counting and Local Encodings}
Both encodings contain $x$, $u$, the bits $q_{j,k}$, their Boolean axioms and the generators~\eqref{eq:parity}; they differ only in how the two cardinality bounds are imposed. The \emph{counting encoding} $E^{\cnt}_{t,b}(H)$ adds Boolean selectors $y_0,\dots,y_t$ and $z_0,\dots,z_b$ with their Boolean axioms, non-Boolean counters $W$ and $U$, and the seven generators
\begin{equation}
\begin{aligned}
&W-\textstyle\sum_{i}x_i,\quad \sum_{\ell=0}^{t}y_\ell-1,\quad W-\sum_{\ell=0}^{t}\ell\, y_\ell,\quad y_0,\\
&U-\textstyle\sum_{j}u_j,\quad \sum_{\varrho=0}^{b}z_\varrho-1,\quad U-\sum_{\varrho=0}^{b}\varrho\, z_\varrho .
\end{aligned}\label{eq:cnt}
\end{equation}
The one-hot selectors restrict $W$ to $\{0,\dots,t\}$ and $U$ to $\{0,\dots,b\}$, and the generator $y_0$ removes $W=0$.

The \emph{local encoding} $E^{\loc}_{t,b}(H)$ has no counters or selectors. It adds the truncated inclusion--exclusion polynomial
\begin{equation}
P_t(x)=\sum_{s=0}^{t}(-1)^s e_s(x),\qquad e_s(x)=\sum_{T\subseteq[n],\,|T|=s}x_T ,\label{eq:Pt}
\end{equation}
where $x_T=\prod_{i\in T}x_i$, and the cardinality obstructions
\begin{equation}
\begin{aligned}
u_R&=\textstyle\prod_{j\in R}u_j, && R\subseteq[m],\ |R|=b+1,\\
x_T&=\textstyle\prod_{i\in T}x_i, && T\subseteq[n],\ |T|=t+1 .
\end{aligned}\label{eq:obs}
\end{equation}
On Boolean points, $P_t(0)=1$ and $P_t(x)=(-1)^t\binom{|S|-1}{t}$ for $S=S(x)\ne\emptyset$, so $P_t$ vanishes exactly when $1\le|S|\le t$, and all $u_R$ vanish exactly when $\sum_ju_j\le b$. The $x_T$ vanish wherever $P_t$ does. Modulo the Boolean axioms, ideals correspond to subsets of the Boolean cube and are radical, so $x_T\in\langle P_t,\,\Bool{x_i}:i\in[n]\rangle$: the $x_T$ change neither the solutions nor the ideal. They are degree cutters in the sense of~\cite{deloera11}, and Section~\ref{sec:local} uses them. None of the obstructions depends on $H$.

\emph{Generator counts.} Counting the Boolean axioms as generators,
\begin{align}
|E^{\cnt}_{t,b}(H)|&=n+2m+\textstyle\sum_jK_j+t+b+9,\label{eq:countcnt}\\
|E^{\loc}_{t,b}(H)|&=n+2m+\textstyle\sum_jK_j+1+\binom{n}{t+1}+\binom{m}{b+1}.\label{eq:countloc}
\end{align}
For the $(3,6)$-regular QC-LDPC matrix with $n=48$ of Section~\ref{sec:verif} and $(t,b)=(3,0)$, the counting encoding has $156$ generators and the local encoding $194{,}749$.

\subsection{Exactness}
An encoding is \emph{infeasible} if its generators have no common zero over $\overline{\Q}$.
\begin{proposition}[Exactness {\cite[Prop.~1]{itw27}}]\label{prop:exact}
Let $1\le t\le n$ and $b\ge 0$. For $E\in\{E^{\cnt}_{t,b}(H),\allowbreak E^{\loc}_{t,b}(H)\}$, every common zero of $E$ over $\overline{\Q}$ has all Boolean variables in $\{0,1\}$, and $x\mapsto S(x)$ is a bijection from the zeros of $E$ onto $\{S\subseteq[n]:1\le|S|\le t,\ \Phi(S)\le b\}$; all auxiliary variables are determined by $x$. Consequently,
\begin{equation}
E^{\cnt}_{t,b}(H)\ \text{infeasible}\iff E^{\loc}_{t,b}(H)\ \text{infeasible}\iff \nexists S:\ 1\le|S|\le t,\ \Phi(S)\le b .\label{eq:equiv}
\end{equation}
\end{proposition}
\begin{IEEEproof}
The axioms $\Bool{v}$ force $v\in\{0,1\}$, and by~\eqref{eq:parity} $u$ and $q$ are then the parity bits and quotient digits of~$x$. In $E^{\cnt}_{t,b}(H)$, $\sum_\ell y_\ell=1$ selects a unique $\ell^*$ with $W=\ell^*=\sum_ix_i$, and $y_0=0$ gives $1\le|S|\le t$; likewise $U=\sum_ju_j=\Phi(S)\le b$ with a unique~$z$. In $E^{\loc}_{t,b}(H)$, $P_t=0$ gives $1\le|S|\le t$, the $x_T$ then vanish, and the $u_R$ give $\Phi(S)\le b$. Conversely, every admissible $S$ extends uniquely to a zero.
\end{IEEEproof}

\subsection{Certificates and Their Degree}
If $E$ is infeasible, the weak Nullstellensatz gives $1\in\langle E\rangle$ over $\overline{\Q}$. Since the generators have rational coefficients, a certificate of fixed degree solves a linear system with rational data, so~\eqref{eq:ns} may be taken over $\Q$, and the minimal degree is the same over $\Q$ and $\overline{\Q}$. We define
\begin{equation}
\degns(E)=\min\ \max_{g\in E}\ \deg(a_gg),\label{eq:degns}
\end{equation}
the minimum over all certificates, with $\degns(E)=\infty$ if $E$ is feasible. This product convention, standard in proof complexity, is used for all degrees below; the multiplier degree $\max_g\deg a_g$ used in~\cite{deloera11} can be smaller by up to $\max_g\deg g$. A \emph{constructed} certificate of degree $D$ shows $\degns(E)\le D$; we say \emph{exact degree} only when a matching lower bound is proved or computed.

For $D\ge0$ let $I_D(E)=\{\sum_{g\in E}a_gg:\ \deg(a_gg)\le D\}$. This is a $\Q$-vector space, $h\cdot I_D(E)\subseteq I_{D+\deg h}(E)$, and $\degns(E)\le D$ if and only if $1\in I_D(E)$; we drop $E$ when it is clear. For Boolean variables, $\ML$ denotes multilinear reduction (replace every $v^2$ by $v$); it satisfies $\ML(fg)=\ML(f\cdot\ML g)$ and does not raise degree.

\begin{lemma}[Dual criterion {\cite[Lemma~4(a)]{itw27}}]\label{lem:dual}
Let $D\ge1$ and let $\Lambda$ be a $\Q$-linear functional on $\Q[\mathbf v]$ with $\Lambda(1)=1$ and $\Lambda(\mu g)=0$ for every $g\in E$ and every monomial $\mu$ with $\deg(\mu g)\le D$. Then $\degns(E)>D$.
\end{lemma}
\begin{IEEEproof}
A certificate of degree at most $D$ may be taken rational. Expanding each $a_g$ into monomials $\mu$, which satisfy $\deg\mu\le D-\deg g$ because $\deg(a_gg)=\deg a_g+\deg g$, and applying $\Lambda$ gives $1=0$.
\end{IEEEproof}

A functional as in Lemma~\ref{lem:dual} is a \emph{design} in the sense of~\cite{beame96,buss98}. For systems whose variables are all Boolean, the following reduction turns membership of $1$ in $I_D$ into finite linear algebra; Section~\ref{sec:verif} uses it for every exact degree computation.

\begin{lemma}[Multilinear reduction]\label{lem:mlred}
Let $E$ consist of the Boolean axioms of all its variables and a set $G$ of further generators. Then $1\in I_D(E)$ if and only if $1$ lies in the $\Q$-span of $\{\ML(\mu g):g\in G,\ \mu$ a multilinear monomial, $\deg\mu+\deg g\le D\}$.
\end{lemma}
\begin{IEEEproof}
For $\deg f\le D$, $f-\ML f$ is a combination of products $\mu'\Bool{v}$ of degree at most $D$, by $v^k=v^{k-2}\Bool{v}+v^{k-1}$; moreover $\ML(\mu g)=\ML(\ML(\mu)\,g)$ with $\deg\ML\mu\le\deg\mu$, and $\ML(\mu'\Bool{v})=0$. A multilinear polynomial in the Boolean ideal vanishes on the cube and is therefore zero, so a certificate of degree at most $D$ exists if and only if $1=\sum_{g\in G}\ML(a_gg)$ for some $a_g$ with $\deg(a_gg)\le D$.
\end{IEEEproof}

\begin{remark}[Characteristic zero]\label{rem:char}
All statements are over $\Q$. The proofs below divide, for example, by the odd integers $c-2Q_j(\varepsilon)$ (the parity inverse in Theorem~\ref{thm:A}), by $w$ and $2$ (Proposition~\ref{prop:B}, Corollary~\ref{cor:B}), by $2$, $3$ and $n-1$ (Proposition~\ref{prop:D}) and by binomial coefficients (Lemma~\ref{lem:A}). Over a field of characteristic $p>0$ the statements can change: for example, $\sum_{v\in\ZM}\eta_v+1=0$ has Boolean solutions modulo $p$ once $M\ge p-1$, so Lemma~\ref{lem:A}, and with it every planted lower bound, is a characteristic-zero statement.
\end{remark}

\subsection{Counter Elimination and a Generic Upper Bound}
\begin{lemma}[Counter elimination]\label{lem:elim-gen}
Let $E'$ be obtained from $E^{\cnt}_{t,b}(H)$ by replacing $W-\sum_ix_i$ with $\sum_\ell\ell y_\ell-\sum_ix_i$ and $U-\sum_ju_j$ with $\sum_\varrho\varrho z_\varrho-\sum_ju_j$, and by deleting $W-\sum_\ell\ell y_\ell$ and $U-\sum_\varrho\varrho z_\varrho$. Then $E'$ has only Boolean variables, has a common zero if and only if $E^{\cnt}_{t,b}(H)$ has one, and $\degns(E')=\degns(E^{\cnt}_{t,b}(H))$.
\end{lemma}
\begin{IEEEproof}
The substitution $W\mapsto\sum_\ell\ell y_\ell$, $U\mapsto\sum_\varrho\varrho z_\varrho$ does not raise degrees. It maps the two deleted generators to $0$ and the two replaced ones to their replacements, and fixes the others; so it maps certificates to certificates of no larger degree. Conversely, the new generators equal $(W-\sum_ix_i)-(W-\sum_\ell\ell y_\ell)$ and $(U-\sum_ju_j)-(U-\sum_\varrho\varrho z_\varrho)$, so every certificate for $E'$ yields one for $E^{\cnt}_{t,b}(H)$ of the same degree. A common zero of either system extends or restricts to one of the other via $W=\sum_\ell\ell y_\ell$ and $U=\sum_\varrho\varrho z_\varrho$.
\end{IEEEproof}

The following bound is folklore (see, e.g.,~\cite{pitassi97}); we include the short proof because the constant matters in Table~\ref{tab:compare} and because Theorem~\ref{thm:cap} reuses it.

\begin{proposition}[Generic upper bound]\label{prop:generic-upper}
Let $E$ consist of the Boolean axioms of $N$ variables and further generators of degree at most $d_G$, without common zero. Then $\degns(E)\le N+d_G$. In particular, if problem~\eqref{eq:problem} is infeasible, then $\degns(E^{\cnt}_{t,b}(H))\le n+m+\sum_jK_j+t+b+3$.
\end{proposition}
\begin{IEEEproof}
Let $G$ be the further generators. The function $F=\sum_{g\in G}g^2$ is positive on $\{0,1\}^N$, since a common zero on the cube would be a common zero of $E$; this uses that $\Q$ is ordered. Let $h_g$ be the multilinear polynomial that equals $g/F$ on $\{0,1\}^N$, so $\deg h_g\le N$. Then $R=1-\sum_gh_gg$ has degree at most $N+d_G$ and vanishes on $\{0,1\}^N$. Reducing $R$ with $v^k=v^{k-2}\Bool v+v^{k-1}$ writes $R=\sum_va_v\Bool v+\ML R$ with $\deg(a_v\Bool v)\le\deg R$, and $\ML R=0$ because it is multilinear and vanishes on the cube. Thus $1=\sum_gh_gg+\sum_va_v\Bool v$ is a certificate of degree at most $N+d_G$. For the second claim apply this to $E'$ of Lemma~\ref{lem:elim-gen}. It has $N=n+m+\sum_jK_j+t+b+2$ Boolean variables and $d_G=1$, and it is infeasible by Proposition~\ref{prop:exact}.
\end{IEEEproof}

\section{Constructive Local Certificates}\label{sec:local}

\subsection{Support Polynomials, Absorption and Parity Inverses}
For $|S|\le t$ let
\begin{equation}
D_S=\sum_{T\supseteq S,\ |T|\le t}(-1)^{|T\setminus S|}\,x_T .\label{eq:DS}
\end{equation}
This is the inclusion--exclusion (M\"obius) expansion of $x_S\prod_{i\notin S}(1-x_i)$ truncated at degree~$t$. The omitted terms are multiples of obstructions $x_T$ with $|T|=t+1$, so on Boolean points of weight at most $t$, $D_S$ is the indicator of $x=\mathbf 1_S$. Since $\sum_{S\subseteq T}(-1)^{|T\setminus S|}=[T=\emptyset]$, we have $\sum_{|S|\le t}D_S=1$ exactly, and $D_\emptyset=P_t$.

\begin{lemma}[Absorption]\label{lem:absorb}
Let $|S|\le t$ and $i\in[n]$.
\begin{enumerate}
\item[(a)] If $i\in S$, then $D_Sx_i-D_S=\Bool{x_i}\sum_T(-1)^{|T\setminus S|}x_{T\setminus\{i\}}$, the sum over $T\supseteq S$ with $|T|\le t$.
\item[(b)] If $i\notin S$, then
\[
D_Sx_i=\Bool{x_i}\sum_{\substack{T\supseteq S\cup\{i\}\\|T|\le t}}(-1)^{|T\setminus S|}x_{T\setminus\{i\}}+\sum_{\substack{T\supseteq S,\ i\notin T\\|T|=t}}(-1)^{|T\setminus S|}x_{T\cup\{i\}} .
\]
\end{enumerate}
All products on the right have degree at most $t+1$.
\end{lemma}
\begin{IEEEproof}
For $T\ni i$, $x_Tx_i=x_T+x_{T\setminus\{i\}}\Bool{x_i}$; this gives~(a). For~(b), split the terms of $D_Sx_i$ according to whether $i\in T$. The terms with $i\in T$ give $\sum(-1)^{|T\setminus S|}x_T$ plus the first sum. A term with $i\notin T$ and $|T|\le t-1$ gives $(-1)^{|T\setminus S|}x_{T\cup\{i\}}$, which cancels the term of $T\cup\{i\}$ in the previous sum: $T\mapsto T\cup\{i\}$ is a bijection onto the sets $T'\supseteq S\cup\{i\}$ with $|T'|\le t$, and the signs differ. The terms with $i\notin T$ and $|T|=t$ remain; they are obstructions $x_{T\cup\{i\}}$ with coefficients $\pm1$.
\end{IEEEproof}

For a check $j$ with $c=|N(j)\cap S|$, applying Lemma~\ref{lem:absorb} to every $i\in N(j)$ gives
\begin{equation}
D_S\,L_j=D_S\,(c-u_j-2Q_j)+\Xi_{S,j},\label{eq:absorb}
\end{equation}
where $\Xi_{S,j}=\sum_{i\in N(j)\cap S}(D_Sx_i-D_S)+\sum_{i\in N(j)\setminus S}D_Sx_i$ is an explicit combination of $x$-Boolean axioms and obstructions whose products have degree at most $t+1$. This is the only place where the $x_T$ enter the certificate; whether comparably low degree is possible without them is open (Section~\ref{sec:disc}).

\emph{Parity inverse.} If $c$ is odd, $c-2Q_j(\varepsilon)$ is an odd integer, hence nonzero, for every $\varepsilon\in\{0,1\}^{K_j}$. Let $\varphi_{j,c}$ be the multilinear polynomial in $q_{j,0},\dots,q_{j,K_j-1}$ interpolating $1/(c-2Q_j)$ on $\{0,1\}^{K_j}$. It has rational coefficients and degree at most $K_j\le K$, and
\begin{equation}
\varphi_{j,c}\,(c-2Q_j)-1=\textstyle\sum_{k}h_k\,\Bool{q_{j,k}},\qquad \deg h_k\le K_j-1 ,\label{eq:inv}
\end{equation}
because the left side vanishes on $\{0,1\}^{K_j}$ and its non-multilinear monomials have the form $q_{j,k}^2\mu$ with $\deg\mu\le K_j-1$.

\subsection{The Sharper Bound}
\begin{theorem}[Constructive local certificate]\label{thm:A}
Let $H\in\F_2^{m\times n}$, $1\le t\le n$ and $b\ge 0$, and suppose that no $S\subseteq[n]$ satisfies $1\le|S|\le t$ and $\Phi(S)\le b$. Then $E^{\loc}_{t,b}(H)$ admits an explicitly constructed rational certificate $1=\sum_ga_g\,g$ with
\begin{equation}
\max_g\deg(a_gg)\le t+b+K+1 .\label{eq:ub}
\end{equation}
It is computed in one pass over the supports of size at most $t$, without Gr\"obner-basis computation. Expanded term by term, before like terms are collected, its multipliers contain at most
\begin{equation}
(b+1)(2d_{\max}+K+3)\,2^{K+t}N_t(n)\label{eq:size}
\end{equation}
monomials in total. Since $2^K\le d_{\max}+1$, this is $O(n^t)$ for fixed $t$, $b$ and $d_{\max}$. Moreover, every certificate for $E^{\loc}_{t,b}(H)$ has degree at least $t$.
\end{theorem}

\begin{IEEEproof}
Since $\sum_SD_S=1$ and $D_\emptyset=P_t$, it suffices to express each $D_S$ with $1\le|S|\le t$ through the generators. Fix such an $S$. Infeasibility gives $\Phi(S)\ge b+1$; let $j_1<\dots<j_{b+1}$ be checks that are odd on $S$, and put $\pi_\ell=\prod_{\ell'<\ell}u_{j_{\ell'}}$ (so $\pi_1=1$) and $u_{R_S}=\prod_{\ell=1}^{b+1}u_{j_\ell}$, an obstruction in~\eqref{eq:obs}. Since $\pi_\ell-\pi_{\ell+1}=\pi_\ell(1-u_{j_\ell})$,
\begin{equation}
D_S=\sum_{\ell=1}^{b}\pi_\ell(1-u_{j_\ell})D_S+\pi_{b+1}D_S .\label{eq:A1}
\end{equation}

\emph{Blocks $\ell\le b$.} Let $j=j_\ell$, $c=|N(j)\cap S|$ (odd) and $\varphi=\varphi_{j,c}$. Combining~\eqref{eq:absorb}, \eqref{eq:inv} and $(1-u_j)(c-u_j-2Q_j)=(1-u_j)(c-2Q_j)+\Bool{u_j}$ gives the exact identity~\cite[(16)]{itw27}
\begin{equation}
\begin{aligned}
\pi_\ell(1-u_j)D_S={}&(1-u_j)\varphi\,\pi_\ell D_S\,L_j-(1-u_j)\varphi\,\pi_\ell\,\Xi_{S,j}\\
&-\pi_\ell D_S(1-u_j)\textstyle\sum_kh_k\Bool{q_{j,k}}-\varphi\,\pi_\ell D_S\,\Bool{u_j}.
\end{aligned}\label{eq:key}
\end{equation}
With $\deg\pi_\ell=\ell-1$, $\deg D_S\le t$ and $\deg\varphi\le K$, the four products have degree at most
\[
\begin{aligned}
&(1{+}K{+}\ell{-}1{+}t)+1,\qquad (1{+}K{+}\ell{-}1)+(t{+}1),\\
&(\ell{-}1{+}t{+}1{+}K{-}1)+2,\qquad (K{+}\ell{-}1{+}t)+2,
\end{aligned}
\]
each equal to $t+\ell+K+1\le t+b+K+1$.

\emph{Last block.} Let $j=j_{b+1}$, $c=|N(j)\cap S|$ (odd) and $\varphi=\varphi_{j,c}$. By~\eqref{eq:absorb}, $(c-2Q_j)D_S=D_SL_j-\Xi_{S,j}+u_jD_S$. Multiplying by $\varphi\,\pi_{b+1}$, using~\eqref{eq:inv}, and noting $\pi_{b+1}u_j=u_{R_S}$,
\begin{equation}
\pi_{b+1}D_S=\varphi\,\pi_{b+1}D_S\,L_j-\varphi\,\pi_{b+1}\,\Xi_{S,j}+\varphi D_S\,u_{R_S}-\pi_{b+1}D_S\textstyle\sum_kh_k\Bool{q_{j,k}}.\label{eq:A2}
\end{equation}
The four products have degree at most $(K{+}b{+}t)+1$, $(K{+}b)+(t{+}1)$, $(K{+}t)+(b{+}1)$ and $(b{+}t{+}K{-}1)+2$, all equal to $t+b+K+1$. The factor $1-u_j$ of~\eqref{eq:key} is not needed here, because the last odd check closes the telescope through the obstruction $u_{R_S}$ itself. If $K_j=0$, then $c=1$, $\varphi=1$ and $h=0$.

Summing~\eqref{eq:A1}--\eqref{eq:A2} over $S\ne\emptyset$ and adding $D_\emptyset=P_t$, of degree $t$, gives the certificate. Finally, the all-zero point annihilates every generator except $P_t$, and $P_t(0)=1$; evaluating any certificate there shows $a_{P_t}\ne0$, so its degree is at least $\deg(a_{P_t}P_t)\ge t$.

\emph{Size.} For $|S|=s$ let $\delta_s$ be the number of sets $T\supseteq S$ with $|T|\le t$; this is the number of terms of $D_S$. The polynomial $\varphi_{j,c}$ has at most $2^{K_j}$ terms. A term $\mu$ of $\varphi_{j,c}$ that contains $q_{j,k}$ contributes the single term $\mu/q_{j,k}$ to $h_k$ in~\eqref{eq:inv}, so the $h_k$ have at most $K_j2^{K_j-1}$ terms together. In Lemma~\ref{lem:absorb}, the terms in~(a) are indexed by the $\delta_s$ sets $T\supseteq S$ with $|T|\le t$, and those in~(b) by two disjoint families of such sets. Hence $\Xi_{S,j}$ is a sum of at most $d_j\delta_s$ terms, each a monomial times a generator.

Expand the products in~\eqref{eq:key} term by term. A block $\ell\le b$ contributes at most $2\cdot2^{K_j}\delta_s$ monomials to the multiplier of $L_j$, $2\cdot2^{K_j}d_j\delta_s$ through $\Xi_{S,j}$, $2\delta_s\cdot K_j2^{K_j-1}$ to those of the $\Bool{q_{j,k}}$, and $2^{K_j}\delta_s$ to that of $\Bool{u_j}$. That is at most $(2d_j+K_j+3)\,2^{K_j}\delta_s$ in total. By~\eqref{eq:A2}, the last block contributes at most $(d_j+K_j/2+2)\,2^{K_j}\delta_s$.

Sum over the $b+1$ blocks and over the supports $S\ne\emptyset$, and add the single term of $D_\emptyset=P_t$. The multipliers then contain at most
\[
1+(b+1)(2d_{\max}+K+3)\,2^K\sum_{s=1}^{t}\binom ns\delta_s
\]
monomials. The sum counts the pairs $\emptyset\ne S\subseteq T$ with $|T|\le t$, so it is at most $(2^t-1)(N_t(n)-1)$, and~\eqref{eq:size} follows. Finally, $K$ is $K_j$ for a check of degree $d_{\max}$, so if $K\ge1$, then $2^{K-1}<\lfloor d_{\max}/2\rfloor+1$ by the definition of $K_j$, and hence $2^K\le d_{\max}+1$. If $K=0$, the inequality is trivial.
\end{IEEEproof}

The bound~\eqref{eq:ub} has four contributions: $t$ for support localization by $D_S$, $b$ for the telescoping product $\pi_{b+1}$, $K$ for the parity inverse, and $1$ for the closing generator, $L_j$ or $u_{R_S}$.

\begin{corollary}[Uniformity in the block length]\label{cor:uniform}
For a family of parity-check matrices with check degrees at most $d$ and fixed $(t,b)$, every infeasible local encoding satisfies
\[
\degns\big(E^{\loc}_{t,b}(H)\big)\le t+b+\big\lceil\log_2(\lfloor d/2\rfloor+1)\big\rceil+1,
\]
uniformly in the block length.
\end{corollary}
This is a statement for fixed $t$ and $b$: if $t$ or $b$ grows with the block length, so does the bound, and Theorem~\ref{thm:A} shows that every local certificate has degree at least $t$.

\subsection{Tightness}
\begin{proposition}[The bound cannot be improved in general]\label{prop:tight}
Let
\[
H_1=\begin{psmallmatrix}0&1&1&0\\1&1&0&1\end{psmallmatrix},\qquad
H_2=\begin{psmallmatrix}0&1&0&1\\0&1&0&1\\0&0&1&1\\1&0&1&1\end{psmallmatrix}.
\]
Both have $K=1$, and
\[
\degns\big(E^{\loc}_{1,0}(H_1)\big)=3,\qquad \degns\big(E^{\loc}_{2,0}(H_2)\big)=4,
\]
which equal $t+b+K+1$ in both cases.
\end{proposition}
\begin{IEEEproof}
Both problems are infeasible by enumeration. Exact computations over $\Q$ (Section~\ref{sec:verif}) give a rational functional as in Lemma~\ref{lem:dual} at degree $D-1$ and a rational certificate at degree $D$, for $D=3$ and $D=4$ respectively. The certificates produced by Theorem~\ref{thm:A} attain these degrees.
\end{IEEEproof}

The bound is not attained on cycle codes (Section~\ref{sec:exact}). In $30$ random infeasible instances with $m\le3$, $n\le5$ and $t+b\le2$, computed exactly over $\Q$, the degree never exceeded $t+b+K+1$; it was below $t+b+1$ in $19$ of them, each containing a check of degree one, and in $30$ further random instances with all check degrees at least two it was always at least $t+b+1$. Degree-one checks are degenerate: if $b=0$ and every variable $i$ lies in a check $j$ of degree one, then $x_i=L_j+u_j\in I_1$ and $1=P_1+\sum_ix_i\in I_1$, so $\degns(E^{\loc}_{1,0}(H))=1=t+b$. Whether $t+b+1$ is a lower bound whenever all check degrees are at least two is open.

\subsection{Exact Degree Two for Cycle and Sparse Regular Codes}\label{sec:exact}
Let $C_n$ ($n\ge3$) be the vertex--edge incidence matrix of the $n$-cycle: check $j\in\Zn$ contains $x_{j-1}$ and $x_j$, so every row and every column has weight two. For $(t,b)=(1,0)$ both encodings are infeasible, since each variable lies in two checks. This example is deliberately simple: it isolates the effect of the encoding. Theorem~\ref{thm:A} gives $\degns(E^{\loc}_{1,0}(C_n))\le3$, where~\cite{itw27} had $4$; the true value is $2$.

\begin{proposition}[Degree two for column-regular matrices]\label{prop:B}
Let every column of $H$ have the same weight $w\ge1$, and let every check have degree at most~$3$. Then $E^{\loc}_{1,0}(H)$ is infeasible and
\[
\degns\big(E^{\loc}_{1,0}(H)\big)\le 2,
\]
independently of $n$. If, in addition, every check has degree at least~$2$, equality holds.
\end{proposition}
\begin{IEEEproof}
For $(t,b)=(1,0)$ the generators are $L_j$, the Boolean axioms, $P_1=1-\sum_ix_i$, the obstructions $u_j$ ($|R|=1$) and $x_ix_k$ ($|T|=2$). Every singleton lies in some check, which then has $c=1$, so the system is infeasible.

If $d_j\le1$, then $K_j=0$ and $s_j=L_j+u_j\in I_1$. If $d_j\in\{2,3\}$, then $K_j=1$ and $Q_j=q_j$, and one checks the polynomial identities
\[
\begin{aligned}
s_j&=(s_j^2-s_j)-4\Bool{q_j}-(s_j+2q_j-2)(L_j+u_j),\\
s_j^2-s_j&=\textstyle\sum_{i\in N(j)}\Bool{x_i}+2\sum_{i<k,\ i,k\in N(j)}x_ix_k .
\end{aligned}
\]
Every product on the right has degree at most~$2$, so $s_j\in I_2$ for all $j$. Since every column has weight $w$, $\sum_js_j=w\sum_ix_i$, and therefore
\[
1=P_1+\frac1w\sum_js_j\in I_2 .
\]
For the lower bound, a degree-one certificate can use only the linear generators $P_1$, $u_j$ and $L_j$, with constant multipliers. If all $d_j\ge2$, then all $K_j=1$, and these generators have the common rational zero $x_i=1/n$, $u_j=0$, $q_{j,0}=d_j/(2n)$; so no such certificate exists.
\end{IEEEproof}

\begin{corollary}\label{cor:B}
For every $n\ge3$, $\degns\big(E^{\loc}_{1,0}(C_n)\big)=2$. An explicit certificate is
\[
1=P_1+\tfrac12\sum_{j\in\Zn}\Big[\,2x_{j-1}x_j+\Bool{x_{j-1}}+\Bool{x_j}-4\Bool{q_j}-(x_{j-1}+x_j+2q_j-2)(L_j+u_j)\Big].
\]
\end{corollary}

\begin{remark}
Call $H$ $(w,d)$-regular if every column has weight $w$ and every check has degree $d$. Proposition~\ref{prop:B} covers all $(w,d)$-regular matrices with $d\in\{2,3\}$, for example the $(2,3)$-regular incidence matrices of the complete graph $\mathsf K_4$ and of the Petersen graph and the $(3,3)$-regular incidence matrix of the Fano plane, for which the local degree is $2$ (also confirmed by exact computation, Section~\ref{sec:verif}). For check degrees at least four, Theorem~\ref{thm:A} gives $K+2$, and whether degree two persists is open.
\end{remark}

\section{A Generalized Planted-Knapsack Theorem}\label{sec:planting}
We plant a knapsack instance into the counting encoding. A few columns are fixed to one, $M$ disjoint sets of columns carry new Boolean variables, and every other variable of $E^{\cnt}_{t,b}(H)$ is mapped to a constant or to an affine function of one new variable. The support count and the syndrome count then become multiples of a single linear form $Y-r$, and a knapsack functional for $Y-r$ pulls back to a functional as in Lemma~\ref{lem:dual}. Since no variable is mapped to a polynomial of degree two or more, no degree is lost.

Throughout, $\eta_v$, $v\in\ZM=\{0,\dots,M-1\}$, are new Boolean variables, unrelated to the selectors $y_\ell$; we write $Y=\sum_{v\in\ZM}\eta_v$, $\eta_A=\prod_{v\in A}\eta_v$ for $A\subseteq\ZM$, and $\binom ra=r(r-1)\cdots(r-a+1)/a!$ for $r\in\Q$ and integers $a\ge0$.

\subsection{The Knapsack Functional with an Arbitrary Target}
\begin{lemma}[Knapsack functional]\label{lem:A}
For $M\ge1$ and $r\in\Q$ let
\[
\Lambda_{M,r}(\eta_A)=\binom{r}{|A|}\Big/\binom{M}{|A|},\qquad A\subseteq\ZM,
\]
extended linearly to multilinear polynomials, and set $\Lambda_{M,r}(f)=\Lambda_{M,r}(\ML f)$ for every polynomial $f$ in $\eta$. Then $\Lambda_{M,r}(1)=1$, $\Lambda_{M,r}(f\,\Bool{\eta_v})=0$ for every $f$ and $v$, and
\[
\Lambda_{M,r}\big(f\cdot(Y-r)\big)=0\qquad\text{whenever }\deg f\le M-1 .
\]
\end{lemma}
\begin{IEEEproof}
$\Lambda_{M,r}(1)=1$ is the case $A=\emptyset$, and $\Lambda_{M,r}(f\Bool{\eta_v})=0$ because $\ML(f\Bool{\eta_v})=0$. For the third claim it suffices to take $f=\eta_A$ with $|A|=a\le M-1$, since $\ML f$ has degree at most $\deg f$. Write $\lambda_a=\binom ra/\binom Ma$. Then $\ML\big(\eta_A(Y-r)\big)=(a-r)\eta_A+\sum_{v\notin A}\eta_{A\cup\{v\}}$, so
\[
\Lambda_{M,r}\big(\eta_A(Y-r)\big)=(a-r)\lambda_a+(M-a)\lambda_{a+1}.
\]
Since $\binom r{a+1}=\binom ra\frac{r-a}{a+1}$ and $\binom M{a+1}=\binom Ma\frac{M-a}{a+1}$, we have $(M-a)\lambda_{a+1}=(r-a)\lambda_a$, and the sum vanishes.
\end{IEEEproof}

For $r=-1$, $\binom{-1}{a}=(-1)^a$ and $\Lambda_{M,-1}(\eta_A)=(-1)^{|A|}/\binom M{|A|}$ is the familiar functional for the knapsack $Y+1=0$. For an unattainable target, $r\notin\{0,\dots,M\}$, the degree limit $M-1$ in Lemma~\ref{lem:A} cannot be raised: $\ML(\eta_{\ZM}(Y-r))=(M-r)\eta_{\ZM}$, so
\begin{equation}
\Lambda_{M,r}\big(\eta_{\ZM}(Y-r)\big)=(M-r)\binom rM ,\label{eq:nonvac}
\end{equation}
which is nonzero if and only if $r\notin\{0,\dots,M\}$. For such $r$ no functional with $\Lambda(1)=1$ vanishes on all multiples of $Y-r$ and of the Boolean axioms of degree at most $M+1$, since $1$ is a combination of such multiples (Proposition~\ref{prop:target}(a)). For $r\in\{0,\dots,M\}$ the target is attained on the cube: $\Lambda_{M,r}(\eta_A)$ is the fraction of the $r$-subsets of $\ZM$ that contain $A$, so $\Lambda_{M,r}$ is the average of the evaluations at the Boolean points with $Y=r$ and vanishes on every multiple of $Y-r$, of any degree.

\subsection{The Knapsack Target}
\begin{proposition}[The knapsack target]\label{prop:target}
For $M\ge1$ and $r\in\Q\setminus\{0,\dots,M\}$ let $\Tgt_{M,r}=\{\Bool{\eta_v}:v\in\ZM\}\cup\{Y-r\}$. Then $\degns(\Tgt_{M,r})=M+1$. More precisely:
\begin{enumerate}
\item[(a)] the polynomial
\[
f_{M,r}=\sum_{A\subseteq\ZM}\frac{(-1)^{|A|}\,|A|!}{\prod_{j=0}^{|A|}(j-r)}\,\eta_A
\]
satisfies $\ML\big(f_{M,r}\,(Y-r)\big)=1$, so that $1=f_{M,r}\,(Y-r)+\sum_vh_v\Bool{\eta_v}$ with every product of degree at most $M+1$;
\item[(b)] if $r\notin\{0,\dots,M-1\}$, then $\Lambda_{M,r}$ is the only linear functional $\Lambda$ on $\Q[\eta]$ with $\Lambda(1)=1$, $\Lambda=\Lambda\circ\ML$, and $\Lambda(f\cdot(Y-r))=0$ whenever $\deg f\le M-1$.
\end{enumerate}
\end{proposition}
\begin{IEEEproof}
(a) On $\{0,1\}^M$, $Y-r$ takes the values $-r,1-r,\dots,M-r$, none of which is zero. By M\"obius inversion, the multilinear interpolant of $1/(Y-r)$ has the coefficient $\sum_{j=0}^{k}\binom kj(-1)^{k-j}/(j-r)$ on $\eta_A$, $|A|=k$. This is the $k$-th forward difference at $0$ of $g_0(x)=1/(x-r)$. With $g_k(x)=(-1)^kk!/\prod_{j=0}^{k}(x+j-r)$,
\[
g_k(x+1)-g_k(x)=\frac{(-1)^kk!\,\big((x-r)-(x+k+1-r)\big)}{\prod_{j=0}^{k+1}(x+j-r)}=g_{k+1}(x),
\]
so the $k$-th difference is $g_k(0)$, the coefficient in $f_{M,r}$. Hence $f_{M,r}(Y-r)-1$ vanishes on the cube and has degree $M+1$. Reducing it with $v^k=v^{k-2}\Bool v+v^{k-1}$ gives the $h_v$, with zero multilinear remainder. Together with Lemma~\ref{lem:A} and Lemma~\ref{lem:dual}, $\degns(\Tgt_{M,r})=M+1$.

(b) Such a $\Lambda$ is determined by the values $\lambda_A=\Lambda(\eta_A)$, $A\subseteq\ZM$. For $|A|=\alpha\le M-1$, the condition with $f=\eta_A$ reads $(\alpha-r)\lambda_A+\sum_{v\notin A}\lambda_{A\cup\{v\}}=0$, and $\alpha-r\ne0$. So $\lambda_A$ is determined by the values on sets of size $\alpha+1$, and by descending induction every $\lambda_A$ is a multiple of $\lambda_{\ZM}$. The solutions form a space of dimension at most one; it contains $\Lambda_{M,r}$, for which $\lambda_\emptyset=1$, so $\Lambda(1)=1$ forces $\Lambda=\Lambda_{M,r}$.
\end{IEEEproof}

For $r=-1$ the coefficients of $f_{M,-1}$ are $(-1)^{|A|}/(|A|+1)$.

\subsection{The Generalized Planting Theorem}
\begin{theorem}[Generalized planting]\label{thm:B}
Let $H\in\F_2^{m\times n}$, $t\ge1$ and $b\ge0$. Let $O\subseteq[n]$, let $M\ge1$, and let $P_0,\dots,P_{M-1}\subseteq[n]\setminus O$ be pairwise disjoint nonempty sets such that
\begin{enumerate}
\item[(R1)] every check meets at most one of the sets $P_v$;
\item[(R2)] $|P_v|=k$ for all $v$;
\item[(R3)] $\lambda_v:=\Phi(O\cup P_v)-\Phi(O)$ is the same number $\lambda$ for all $v$;
\item[(R4)] there are integers $c\in\{1,\dots,t\}$ and $\beta\in\{0,\dots,b\}$ with
\[
k\,\big(\Phi(O)-\beta\big)=\lambda\,\big(|O|-c\big).
\]
\end{enumerate}
Then $\degns\big(E^{\cnt}_{t,b}(H)\big)\ge M+1$.
\end{theorem}

The theorem does not assume that problem~\eqref{eq:problem} is infeasible; if it is feasible, $\degns=\infty$ and the bound is trivial. Put
\begin{equation}
r=\frac{c-|O|}{k},\qquad T=Y-r .\label{eq:target}
\end{equation}
By (R1), the planted sets change the parities of disjoint sets of checks, so $\Phi\big(O\cup\bigcup_{v\in A}P_v\big)=\Phi(O)+\lambda|A|$ for every $A\subseteq\ZM$. If $r\in\{0,\dots,M\}$ and $|A|=r$, this support has size $|O|+kr=c\in[1,t]$ and syndrome weight $\Phi(O)+\lambda r=\beta\le b$ by (R4), so problem~\eqref{eq:problem} is feasible. Hence, whenever the bound is not trivial, $r\notin\{0,\dots,M\}$.

\begin{IEEEproof}
We define a ring homomorphism $\tau$ from the polynomial ring of $E^{\cnt}_{t,b}(H)$ to $\Q[\eta_0,\dots,\eta_{M-1}]$ by its values on the variables, determine the image of every generator, and pull back $\Lambda_{M,r}$.

\emph{Step 1 (support variables).} Put $\tau(x_i)=1$ for $i\in O$, $\tau(x_i)=\eta_v$ for $i\in P_v$, and $\tau(x_i)=0$ otherwise. This is well defined because the $P_v$ are pairwise disjoint and disjoint from $O$.

\emph{Step 2 (parity variables).} Fix a check $j$ and let $a_0=c_j(O)$. If $N(j)$ meets some $P_v$, which is unique by (R1), let $a_1=a_0+|N(j)\cap P_v|$ and $\eta=\eta_v$; otherwise let $a_1=a_0$ and $\eta=0$. Then $\tau(s_j)=a_0+(a_1-a_0)\eta$. For a function $\phi$ on $\{a_0,a_1\}$ write $\phi^\tau=\phi(a_0)+(\phi(a_1)-\phi(a_0))\eta$, and set
\[
\tau(u_j)=(a\bmod 2)^\tau,\qquad \tau(q_{j,k'})=\big(\text{bit }k'\text{ of }\lfloor a/2\rfloor\big)^\tau .
\]
The bits exist because $\lfloor a_1/2\rfloor\le\lfloor d_j/2\rfloor\le2^{K_j}-1$. The map $\phi\mapsto\phi^\tau$ is linear, and $a=(a\bmod2)+2\lfloor a/2\rfloor$ for every integer $a$, so $\tau(L_j)=\big(a-(a\bmod2)-2\lfloor a/2\rfloor\big)^\tau=0$.

\emph{Step 3 (Boolean axioms of $x$, $u$, $q$).} Every value $\phi^\tau$ above lies in $\{0,1,\eta,1-\eta\}$ because $\phi$ takes values in $\{0,1\}$, and $\Bool{1-\eta}=\Bool{\eta}$. So every Boolean axiom of an $x$, $u$ or $q$ variable maps to $0$ or to some $\Bool{\eta_v}$.

\emph{Step 4 (the two sums).} By (R2), $\tau(\sum_ix_i)=|O|+kY$. A check that meets no planted set keeps the parity $c_j(O)\bmod2$. The checks that meet $P_v$ contribute $\sum_j\big(c_j(O\cup P_v)\bmod 2-c_j(O)\bmod2\big)\eta_v=\lambda_v\eta_v$ in addition, by the definition of $\lambda_v$. By (R3),
\[
\tau\Big(\sum_ju_j\Big)=\Phi(O)+\lambda Y .
\]

\emph{Step 5 (selectors and counters).} Put $\tau(y_c)=1$, $\tau(y_\ell)=0$ for $\ell\ne c$, $\tau(W)=c$, $\tau(z_\beta)=1$, $\tau(z_\varrho)=0$ for $\varrho\ne\beta$, and $\tau(U)=\beta$.

\emph{Step 6 (images of the counting generators).} By Step~4, and since (R4) gives $\beta-\Phi(O)=\lambda(c-|O|)/k=\lambda r$,
\begin{equation}
\begin{aligned}
\tau\Big(W-\sum_ix_i\Big)&=c-|O|-kY=-k\,T,\\
\tau\Big(U-\sum_ju_j\Big)&=\beta-\Phi(O)-\lambda Y=-\lambda\,T.
\end{aligned}\label{eq:twoimages}
\end{equation}
The generators $\sum_\ell y_\ell-1$, $W-\sum_\ell\ell y_\ell$, $\sum_\varrho z_\varrho-1$ and $U-\sum_\varrho\varrho z_\varrho$ map to $0$; $y_0$ maps to $0$ because $c\ge1$; and every Boolean axiom of a selector maps to $0$.

\emph{Step 7 (degree preservation).} Every variable, including $W$ and $U$, is mapped to a polynomial of degree at most one, so $\deg\tau(\mu)\le\deg\mu$ for every monomial $\mu$ (where $\tau(\mu)$ may be $0$).

\emph{Step 8 (the functional).} Let $\Lambda=\Lambda_{M,r}\circ\tau$, so $\Lambda(1)=1$. Let $g$ be a generator and $\mu$ a monomial with $\deg(\mu g)\le M$. If $\tau(g)=0$, then $\Lambda(\mu g)=0$. If $\tau(g)=\Bool{\eta_v}$, then $\Lambda(\mu g)=\Lambda_{M,r}(\tau(\mu)\Bool{\eta_v})=0$ by Lemma~\ref{lem:A}. Otherwise $g$ is one of the two generators in~\eqref{eq:twoimages}. Both have degree one, so $\deg\tau(\mu)\le\deg\mu\le M-1$ by Step~7, and $\Lambda(\mu g)=-\gamma\,\Lambda_{M,r}\big(\tau(\mu)\,T\big)=0$ by Lemma~\ref{lem:A}, with $\gamma\in\{k,\lambda\}$. Lemma~\ref{lem:dual} with $D=M$ gives $\degns(E^{\cnt}_{t,b}(H))>M$.
\end{IEEEproof}

Only two generators of degree one carry the target, and the knapsack functional annihilates every multiple of $T$ by a polynomial of degree at most $M-1$. This is why the bound reaches degree $M$ and no further: for $\mu=\prod_vx_{i_v}$ with $i_v\in P_v$, \eqref{eq:nonvac} gives
\begin{equation}
\Lambda\Big(\mu\,\big(W-\textstyle\sum_ix_i\big)\Big)=-k\,(M-r)\binom rM\ne0,\label{eq:nonvac-planted}
\end{equation}
a product of degree $M+1$. Section~\ref{sec:limits} shows that no pullback of this kind proves more.

\begin{remark}[A selector can carry the target]\label{rem:selector}
One can also let a selector carry the target: $\tau(y_c)=1+\frac kcT$, $\tau(W)=c+kT$, $\tau(U)=\beta+\lambda T$, and, if $\beta\ge1$, $\tau(z_\beta)=1+\frac\lambda\beta T$. Then $W-\sum_ix_i$ and $U-\sum_ju_j$ map to $0$, the other selector generators map to $0$ or to constant multiples of $T$, and the Boolean axioms of $y_c$ and $z_\beta$ map to $T$ times linear polynomials; since these axioms have degree two, the same argument applies. Both substitutions cover exactly the same instances, since (R1)--(R4) are unchanged. The constant selectors used above keep every Boolean axiom inside the Boolean ideal of the target, which Section~\ref{sec:limits} uses.
\end{remark}

\subsection{Single Planted Columns}
For $k=1$ and $c=|O|-1$, the target is $Y+1$, and Theorem~\ref{thm:B} specializes to the following two statements. We write $i_v$ for the column of $P_v=\{i_v\}$.

\begin{corollary}[Planted knapsack]\label{cor:planted}
Let $H\in\F_2^{m\times n}$, $t\ge1$ and $b\ge0$, and assume:
\begin{enumerate}
\item[(P1)] $O\subseteq[n]$ and $2\le|O|\le t+1$;
\item[(P2)] $i_0,\dots,i_{M-1}\in[n]\setminus O$ are distinct columns, $M\ge1$, all of the same weight $w_*$;
\item[(P3)] $C(i_v)\cap C(i_{v'})=\emptyset$ for all $v\ne v'$;
\item[(P4)] $C(i_v)\cap C(O)=\emptyset$ for all $v$;
\item[(P5)] $\beta:=\Phi(O)-w_*$ satisfies $0\le\beta\le b$.
\end{enumerate}
Then $\degns\big(E^{\cnt}_{t,b}(H)\big)\ge M+1$.
\end{corollary}
\begin{IEEEproof}
Take $P_v=\{i_v\}$, $k=1$ and $c=|O|-1\in\{1,\dots,t\}$ by (P1). (R1) is (P3) and (R2) holds. By (P4), adding $i_v$ to $O$ makes all $w_*$ checks of $i_v$ odd, so $\lambda_v=w_*$, which is (R3). Finally $\Phi(O)-\beta=w_*=\lambda(|O|-c)$, which is (R4).
\end{IEEEproof}

\begin{corollary}[Relaxed planting]\label{cor:planted-relaxed}
Let $H$, $t$, $b$ and $O$ satisfy (P1), and let $i_0,\dots,i_{M-1}\in[n]\setminus O$ be distinct columns, $M\ge1$, satisfying (P3). For each $v$ let $\nu_v=\#\{j\in C(i_v):c_j(O)\text{ odd}\}$ and $\lambda_v=|C(i_v)|-2\nu_v$. If all $\lambda_v$ equal a common value $\lambda$ and $\beta:=\Phi(O)-\lambda$ satisfies $0\le\beta\le b$, then $\degns(E^{\cnt}_{t,b}(H))\ge M+1$.
\end{corollary}
\begin{IEEEproof}
Adding $i_v$ to $O$ flips the parity of the checks of $C(i_v)$: $\nu_v$ odd checks become even and $|C(i_v)|-\nu_v$ even checks become odd, so $\Phi(O\cup\{i_v\})-\Phi(O)=\lambda_v$. Theorem~\ref{thm:B} applies with $k=1$, $c=|O|-1$ and $\lambda(|O|-c)=\lambda=\Phi(O)-\beta$.
\end{IEEEproof}

Under (P2) and (P4), $\nu_v=0$ and $\lambda=w_*$, so Corollary~\ref{cor:planted} is the case $\nu_v=0$ of Corollary~\ref{cor:planted-relaxed}. In both, $\tau(u_j)\in\{\eta_v,1-\eta_v\}$ on the planted checks and the substitution of Theorem~\ref{thm:B} is the one with constant selectors $\tau(y_{|O|-1})=1$, $\tau(W)=|O|-1$, $\tau(z_\beta)=1$ and $\tau(U)=\beta$.

\begin{remark}[Hypotheses of the single-column plantings]\label{rem:planted-hyp}
(a)~\emph{Characteristic zero} enters only through Lemma~\ref{lem:A}, which divides by $\binom M\alpha$; the map $\tau$ has integer coefficients, and Lemma~\ref{lem:dual} holds over every field (Remark~\ref{rem:char}).
(b)~\emph{Endpoints.} For $M=1$ the bound is $2$; the proof then reduces to the rational point obtained from $\tau$ with $\eta_0=-1$, which is a common zero of all generators of degree one. For $t=1$, (P1) forces $|O|=2$; for $b=0$, (P5) forces $\Phi(O)=w_*$. If $w_*=0$, the planted columns are zero columns, the problem is feasible and the statement is vacuous.
(c)~\emph{Necessity within single-column plantings with target $Y+1$.} $|O|\ge2$ because $\tau(y_0)$ must vanish while $\tau(W)=|O|-1$; $|O|\le t+1$ because $y_{|O|-1}$ must exist; (P3), because two planted columns in one check would make $\tau(u_j)$ the quadratic parity $\eta_v+\eta_{v'}-2\eta_v\eta_{v'}$, which costs a factor $\frac12$ (Theorem~\ref{thm:red}); and (P5), because $\tau(U-\sum_ju_j)$ must be a multiple of $Y+1$. Theorem~\ref{thm:B} removes the first two through the general target (Section~\ref{sec:removes}).
\end{remark}

\subsection{What the General Target and Even Planted Sets Remove}\label{sec:removes}
The single-column plantings with target $Y+1$ have two restrictions: $|O|\le t+1$ and a parity rule. They are removed by different ingredients of Theorem~\ref{thm:B}.

\emph{The condition $|O|\le t+1$.} With $k=1$ and $r=-1$, the selector value is forced to be $c=|O|+kr=|O|-1$, and $c\in\{1,\dots,t\}$ gives $2\le|O|\le t+1$. With a general target any $c\in\{1,\dots,t\}$ can be used, with $r=(c-|O|)/k$, so $|O|$ is unrestricted. Theorem~\ref{thm:H} uses $|O|=d-s$ with $t$ as small as $d-2s$.

\emph{The parity rule.}
\begin{lemma}[Parity rule]\label{lem:parity}
For every $O\subseteq[n]$, $\Phi(O)=\sum_{o\in O}|C(o)|-2\sum_j\lfloor c_j(O)/2\rfloor$. Consequently, in Corollary~\ref{cor:planted-relaxed}, $\beta\equiv\sum_{o\in O}|C(o)|-|C(i_v)|\pmod2$ for every $v$. If all columns have weight $w$, then $\beta\equiv w(|O|-1)\pmod2$; in particular, for $t=1$ and odd $w$, $\beta$ is odd, so $b\ge1$ is necessary for Corollaries~\ref{cor:planted} and~\ref{cor:planted-relaxed}.
\end{lemma}
\begin{IEEEproof}
$\sum_jc_j(O)=\sum_{o\in O}|C(o)|$, and $c-[c\text{ odd}]=2\lfloor c/2\rfloor$ for every integer $c\ge0$. Since $\lambda_v\equiv|C(i_v)|\pmod2$, the congruence for $\beta$ follows. For $t=1$, (P1) forces $|O|=2$.
\end{IEEEproof}

\begin{lemma}[Parity of (R4)]\label{lem:G}
Let every column of $H$ have weight $w$, and let the planted sets in Theorem~\ref{thm:B} have odd size $k$. Then (R4) implies $\beta\equiv wc\pmod 2$. In particular, if $w$ is odd and $t=1$, then $\beta$ is odd, so $b\ge1$.
\end{lemma}
\begin{IEEEproof}
Adding a column $i\notin A$ to a set $A$ changes the parity of exactly the checks in $C(i)$, so $\Phi(A\cup\{i\})\equiv\Phi(A)+w\pmod 2$. Hence $\Phi(A)\equiv w|A|$ for every $A$; in particular $\Phi(O)\equiv w|O|$ and $\lambda\equiv wk\equiv w$. Reducing (R4) modulo $2$, with $k$ odd, gives $w|O|-\beta\equiv w(|O|-c)$, that is, $\beta\equiv wc$.
\end{IEEEproof}

So planted sets of odd size are subject to the parity rule whatever the selectors do. For even $k$ both sides of (R4) are even and the congruence says nothing; Corollary~\ref{cor:E} attains $\beta=0$ at $t=1$ for every $w$ with $k=2$. For $t\ge2$ an even $c$ also suffices, and Theorem~\ref{thm:H} has instances with $k=1$ and $c=2$. In short: \emph{the condition $|O|\le t+1$ and the parity rule constrain only the single-column plantings with target $Y+1$; Theorem~\ref{thm:B} removes the first through general targets, and the second through planted sets of even size or through an even $c$ when $t\ge2$.}

\section{Linear Separations for Cycle, Regular and QC-LDPC Codes}\label{sec:sep}
This section applies the single-column plantings of Section~\ref{sec:planting}. Each family below consists of infeasible instances whose infeasibility is easy to see combinatorially (from the column weight, the girth or a short enumeration); the lower bounds concern the length of an algebraic proof in the counting encoding, not the difficulty of the decision problem.

\subsection{Tanner-Graph Conditions}
Two columns share a check exactly when they are at distance two in the Tanner graph $T(H)$. Conditions (P3) and~(P4) say that the planted columns are pairwise at distance at least four, and at distance at least four from $O$.

\begin{lemma}[Structural conditions]\label{lem:structure}
Let $H\in\F_2^{m\times n}$ have check degrees at most $d\ge1$.
\begin{enumerate}
\item[(a)] If $O=\{o_1,o_2\}$ and the two columns share exactly $s$ checks, then $\Phi(O)=|C(o_1)|+|C(o_2)|-2s$.
\item[(b)] If $T(H)$ has girth at least $6$ and contains a $6$-cycle through the columns $o_1,o_2,o_3$, then each pair of them shares exactly one check, the three shared checks are distinct, and for $O=\{o_1,o_2,o_3\}$, $|C(O)|=\sum_k|C(o_k)|-3$ and $\Phi(O)=\sum_k|C(o_k)|-6$.
\item[(c)] For $O\subseteq[n]$ and $w_*\ge1$, let $A$ be the set of columns of weight $w_*$ outside $O$ that share no check with $O$, and let $n_{w_*}$ be the number of columns of weight $w_*$. Then $|A|\ge n_{w_*}-|O|-|C(O)|(d-1)$, and $A$ contains
\[
M\ge\Big\lceil\frac{|A|}{1+w_*(d-1)}\Big\rceil
\]
columns with pairwise disjoint check sets. These columns satisfy (P2)--(P4).
\end{enumerate}
\end{lemma}
\begin{IEEEproof}
(a) and (b) follow from Lemma~\ref{lem:parity}. In~(a), $c_j(O)=2$ on the $s$ shared checks and $c_j(O)\le1$ elsewhere. In~(b), two of the columns cannot share a second check, since that would close a $4$-cycle. So the shared checks are the three checks $j_{12},j_{23},j_{31}$ of the $6$-cycle, which are distinct. No check contains all three columns: it would be shared by every pair, hence equal to $j_{12}$, $j_{23}$ and $j_{31}$. Therefore $c_j(O)=2$ on these three checks and $c_j(O)\le1$ elsewhere, and both formulas follow. (c) A check $j\in C(O)$ contains at least one column of $O$, hence at most $d-1$ columns outside $O$. So at most $|C(O)|(d-1)$ columns outside $O$ share a check with $O$, which gives the bound on $|A|$. A column of $A$ shares a check with at most $w_*(d-1)$ other columns. Choosing columns of $A$ greedily, and discarding after each choice the columns that share a check with it, gives the stated number of columns with pairwise disjoint check sets.
\end{IEEEproof}

\begin{corollary}[Linear lower bounds]\label{cor:linear-general}
Fix $t\ge1$, $b\ge0$, $d$, $w_*\ge1$, $s_0$ and $\epsilon>0$. Let $(H_n)$ be parity-check matrices with $n$ columns and check degrees at most $d$, each with a set $O_n$ satisfying (P1), $|C(O_n)|\le s_0$ and $0\le\Phi(O_n)-w_*\le b$, and each with at least $\epsilon n$ columns of weight $w_*$. Then
\[
\degns\big(E^{\cnt}_{t,b}(H_n)\big)\ge1+\frac{\epsilon n-(t+1)-s_0(d-1)}{1+w_*(d-1)}=\Omega(n).
\]
\end{corollary}
\begin{IEEEproof}
Lemma~\ref{lem:structure}(c) and Corollary~\ref{cor:planted}.
\end{IEEEproof}

\begin{corollary}[Regular codes of girth at least six]\label{cor:regular}
Let $H$ be $(w,d)$-regular with $w,d\ge2$ and Tanner graph of girth at least $6$. For $w-2\le b\le w-1$, problem~\eqref{eq:problem} with $t=1$ is infeasible and
\[
\begin{aligned}
\degns\big(E^{\cnt}_{1,b}(H)\big)&\ge1+\max\{0,M_0\},\\
\degns\big(E^{\loc}_{1,b}(H)\big)&\le b+K+2 ,
\end{aligned}
\]
where $M_0=\big\lceil\frac{n-2-(2w-1)(d-1)}{1+w(d-1)}\big\rceil$.
\end{corollary}
\begin{IEEEproof}
Every singleton has $\Phi=w>b$. Two columns of a common check share exactly one check by the girth condition. By Lemma~\ref{lem:structure}(a) they form a set $O$ with $\Phi(O)=2w-2$, $\beta=w-2\in[0,b]$ and $|C(O)|=2w-1$. Lemma~\ref{lem:structure}(c) with $n_w=n$ and Corollary~\ref{cor:planted} give the first bound (it is trivial when the maximum is $0$). Theorem~\ref{thm:A} with $t=1$ gives the second.
\end{IEEEproof}

For fixed $(w,d)$ the lower bound is linear in $n$, and by Proposition~\ref{prop:generic-upper} the counting degree is $O(n)$, so this is a $\Theta(n)$-versus-$O(1)$ separation at $(t,b)=(1,b)$, $w-2\le b\le w-1$. The range of $b$ is forced by Lemma~\ref{lem:parity}; Section~\ref{sec:odd} reaches every $b\le w-1$, including $b=0$, with planted pairs.

\subsection{Cycle Codes and the Compact System $\OPP_n$}
Recall that $C_n$ is the incidence matrix of the $n$-cycle (Section~\ref{sec:exact}).
\begin{corollary}[Cycle codes]\label{cor:lin-cycle}
Let $n\ge5$.
\begin{enumerate}
\item[(a)] For $1\le t\le n-1$ and $b\in\{0,1\}$, problem~\eqref{eq:problem} is infeasible on $C_n$ and $\degns\big(E^{\cnt}_{t,b}(C_n)\big)\ge\lfloor(n-1)/2\rfloor$.
\item[(b)] For $t=1$ and $b\in\{0,1\}$, the largest $M$ admitted by Corollaries~\ref{cor:planted} and~\ref{cor:planted-relaxed} is $\lfloor(n-3)/2\rfloor$, so $\lfloor(n-1)/2\rfloor$ is the strongest bound they give.
\item[(c)] $\lfloor(n-1)/2\rfloor\le\degns\big(E^{\cnt}_{1,0}(C_n)\big)\le3n+4$ and $\degns\big(E^{\loc}_{1,0}(C_n)\big)=2$.
\end{enumerate}
\end{corollary}
\begin{IEEEproof}
Here $C(i)=\{i,i+1\}$, and $\Phi(S)\equiv2|S|\equiv0\pmod2$ for every $S$.

(a) The kernel of $C_n$ is $\{0,\mathbf 1\}$, so no nonempty support of size at most $n-1$ has $\Phi=0$, and by parity none has $\Phi=1$. Take $O=\{p,p+1\}$, two columns sharing the check $p+1$. By Lemma~\ref{lem:structure}(a), $\Phi(O)=2=w_*$, so $\beta=0$; and $|O|=2\le t+1$. A column $i$ shares a check with $O$ exactly when $i\in\{p-1,\dots,p+2\}$, and two columns share a check exactly when they are cyclically adjacent. The other $n-4$ columns, $p+3,\dots,p+n-2$, are consecutive and contain $\lceil(n-4)/2\rceil=\lfloor(n-3)/2\rfloor$ pairwise nonadjacent columns. Corollary~\ref{cor:planted} gives $\degns\ge\lfloor(n-3)/2\rfloor+1=\lfloor(n-1)/2\rfloor$.

(b) For $t=1$, $|O|=2$. Every $\lambda_v=2-2\nu_v$ is even, and so is $\Phi(O)$, so $\beta\in[0,1]$ forces $\beta=0$ and $\lambda=\Phi(O)\le2$. A pair sharing no check has $\Phi(O)=4$, so $O=\{p,p+1\}$, $\lambda=2$ and every $\nu_v=0$. The odd checks of $O$ are $p$ and $p+2$, and check $p+1$ contains only the columns of $O$. So $\nu_v=0$ means that $i_v$ shares no check with $O$, and (P3) means that the planted columns are pairwise nonadjacent. Hence $M\le\lfloor(n-3)/2\rfloor$, as in~(a).

(c) The lower bound is~(a), and the local value is Corollary~\ref{cor:B}. The upper bound is Proposition~\ref{prop:generic-upper} with $\sum_jK_j=n$, $t=1$ and $b=0$.
\end{IEEEproof}

Plantings of Theorem~\ref{thm:B} with larger planted sets do not improve~(b) in general: an exhaustive search over all plantings with $|P_v|\le3$ for $n\le9$ and $|P_v|\le2$ for $10\le n\le12$ (Section~S-III of the supplementary material) finds the maximum $M=\lfloor(n-3)/2\rfloor$ at $t=1$, $b\in\{0,1\}$, with one exception: at $(n,b)=(6,1)$, $O=\emptyset$ with the two planted pairs $\{0,1\}$ and $\{3,4\}$, $k=2$, $\lambda=2$, $c=1$ and $\beta=1$ give $M=2$, so $\degns(E^{\cnt}_{1,1}(C_6))\ge3$.

The counting degree on $C_n$ is related to a compact Boolean system. Let $p_1=\sum_{i\in\Zn}x_i$, $p_2=\sum_{i\in\Zn}x_{i-1}x_i$ and
\begin{equation}
\OPP_n=\{\,\Bool{x_i}\ (i\in\Zn),\ \ p_1-1,\ \ p_2-1\,\}.\label{eq:opp}
\end{equation}
At $(t,b)=(1,0)$ the counting generators~\eqref{eq:cnt} force $p_1=W=1$ and $U=0$, and eliminating the parity variables turns $U-\sum_ju_j=0$ into $p_2=p_1$.

\begin{theorem}[Cycle reduction {\cite[Thm.~6]{itw27}}]\label{thm:red}
For $n\ge3$, $\degns(E^{\cnt}_{1,0}(C_n))\ge\big\lceil\tfrac12\degns(\OPP_n)\big\rceil$.
\end{theorem}
\begin{IEEEproof}
Let $\psi$ fix $x$ and set
\begin{equation}
u_j\mapsto x_{j-1}+x_j-2x_{j-1}x_j,\qquad q_j\mapsto x_{j-1}x_j,\qquad y_1,W,z_0\mapsto1,\qquad y_0,U\mapsto0 .\label{eq:subst}
\end{equation}
Then $\psi(L_j)=0$, $\psi$ fixes $\Bool{x_i}$, the remaining selector generators and their Boolean axioms map to~$0$, and
\[
\begin{aligned}
W-\textstyle\sum_ix_i&\mapsto-(p_1-1),\\
U-\textstyle\sum_ju_j&\mapsto-2(p_1-1)+2(p_2-1),\\
\Bool{q_j}&\mapsto x_j^2\Bool{x_{j-1}}+x_{j-1}\Bool{x_j},\\
\Bool{u_j}&\mapsto(1-4x_j+4x_j^2)\Bool{x_{j-1}}+\Bool{x_j}.
\end{aligned}
\]
Each image is a combination of $\OPP_n$ generators whose products have degree at most $2\deg g$, and $\psi$ at most doubles the degree of every monomial. A certificate of degree $D$ therefore maps to an $\OPP_n$ certificate of degree at most $2D$, so $\degns(\OPP_n)\le2\degns(E^{\cnt}_{1,0}(C_n))$, and the ceiling follows because degrees are integers.
\end{IEEEproof}

\begin{lemma}[Pull-back for $\OPP_n$]\label{lem:C2}
Let $\tau$ substitute each $x_i$, $i\in\Zn$, by $\tau_i\in\{0,1\}\cup\{\eta_v:v\in\ZM\}$. Suppose that
\begin{enumerate}
\item[(a)] $\tau(p_1-1)=\gamma\,(Y+1)$ for some $\gamma\in\Q$, and
\item[(b)] $\ML\,\tau(p_2-1)=\ML\big((Y+1)\,h\big)$ for some polynomial $h$ with $\deg h\le1$.
\end{enumerate}
Then $\degns(\OPP_n)\ge M+1$.
\end{lemma}
\begin{IEEEproof}
Set $\Lambda(F)=\Lambda_{M,-1}(\tau(F))$ for polynomials $F$ in $x$; $\Lambda$ is $\Q$-linear with $\Lambda(1)=1$. Three facts are used. (i)~\emph{$\tau$ does not raise degree}: each $\tau_i$ is a constant or a single variable, so $\deg\tau(\mu)\le\deg\mu$ for every monomial $\mu$. (ii)~\emph{$\ML$ is multiplicative modulo the Boolean ideal}, $\ML(fg)=\ML(f\cdot\ML g)$, so hypothesis~(b) can be used inside a product. (iii)~\emph{Monomials suffice}, as in Lemma~\ref{lem:dual} with $D=M$.

Now check each generator. \emph{Boolean axioms}: $\tau(\Bool{x_i})$ is $0$ or $\Bool{\eta_v}$, so $\Lambda(\mu\Bool{x_i})=0$ for every $\mu$. \emph{$g=p_1-1$}, $\deg\mu\le M-1$: $\Lambda(\mu g)=\gamma\Lambda_{M,-1}\big(\tau(\mu)(Y+1)\big)=0$ by Lemma~\ref{lem:A}, since $\deg\tau(\mu)\le M-1$ by~(i). \emph{$g=p_2-1$}, $\deg\mu\le M-2$: by~(ii) and~(b), $\Lambda(\mu g)=\Lambda_{M,-1}\big(\ML(\tau(\mu)\,h\,(Y+1))\big)$, which is $0$ by Lemma~\ref{lem:A} because $\deg(\tau(\mu)h)\le M-1$. Lemma~\ref{lem:dual} completes the proof.
\end{IEEEproof}

The argument needs $\tau$ to have degree at most one: a substitution of degree $\delta$ maps a certificate of degree $D$ to one of degree at most $\delta D$, so the same argument would exclude only $D\le M/\delta$, which is the source of the factor $\tfrac12$ in Theorem~\ref{thm:red}. The factorization in~(b) costs no degree, since the multiplier passed to Lemma~\ref{lem:A} has degree $\deg\mu+1$, as for $p_2-1$ itself.

\begin{corollary}[Linear growth of the compact system]\label{cor:lin-opp}
For every $n\ge5$, $\degns(\OPP_n)\ge\lfloor(n-1)/2\rfloor$.
\end{corollary}
\begin{IEEEproof}
Let $M=\lfloor(n-3)/2\rfloor$, so $2M+3\le n$, and label the positions of $\Zn$ by $\eta_0,0,\eta_1,0,\dots,\eta_{M-1},0,1,1,0$, followed by zeros; position $n-1$ carries a $0$. Then $\tau(p_1-1)=Y+1$, and the only cyclically adjacent pair with both labels nonzero is the pair $(1,1)$, so $\tau(p_2-1)=0$. This is Lemma~\ref{lem:C2} with $\gamma=1$ and $h=0$.
\end{IEEEproof}

Exact computations over $\Q$ (Section~\ref{sec:verif}) give $\degns(\OPP_n)=\lceil n/2\rceil+1$ for $4\le n\le12$ (\cite{itw27} reported $4\le n\le10$).

\begin{conjecture}\label{conj}
$\degns(\OPP_n)=\lceil n/2\rceil+1$ for all $n\ge4$.
\end{conjecture}
Corollary~\ref{cor:lin-opp} establishes the linear growth of $\degns(\OPP_n)$, which was the asymptotic content of the conjecture in~\cite{itw27}; the exact formula remains open, and the lower bound $\lfloor(n-1)/2\rfloor$ is exactly $2$ below $\lceil n/2\rceil+1$ for $5\le n\le12$. An earlier embedding, which places all $\binom M2$ pairs of knapsack variables at adjacent positions of $\Zn$ by means of Walecki's decomposition of the complete graph $\mathsf K_M$ into Hamiltonian paths~\cite{alspach08}, gives only $\degns(\OPP_n)>\sqrt{2n}-3$ and hence $\degns(E^{\cnt}_{1,0}(C_n))>(\sqrt{2n}-3)/2$. It is superseded by Corollaries~\ref{cor:lin-cycle} and~\ref{cor:lin-opp}; its statement and proof are given in Section~S-I of the supplementary material.

\subsection{Quasi-Cyclic Codes}
Let $H$ be an array of $Z\times Z$ blocks, each zero or a circulant permutation matrix $P^e$, where $(P^e)_{a,\kappa}=1$ if and only if $a\equiv\kappa+e\pmod Z$. Column $\kappa$ of column block $\gamma$ is denoted $(\gamma,\kappa)$, and check $a$ of row block $\rho$ is denoted $(\rho,a)$. The exponent of a nonzero block $(\rho,\gamma)$ is $e_{\rho,\gamma}$.

\begin{lemma}[Circulant structure]\label{lem:qc}
(a)~Distinct columns of the same column block have disjoint check sets. (b)~Every check meets every column block in at most one column. (c)~If columns $(\gamma_1,\kappa_1)$ and $(\gamma_2,\kappa_2)$ share two checks, then $\gamma_1\ne\gamma_2$, the two checks lie in distinct row blocks $\rho_1\ne\rho_2$, the four blocks $(\rho_i,\gamma_l)$ are nonzero, and $e_{\rho_1,\gamma_1}-e_{\rho_1,\gamma_2}+e_{\rho_2,\gamma_2}-e_{\rho_2,\gamma_1}\equiv0\pmod Z$.
\end{lemma}
\begin{IEEEproof}
In every nonzero row block $\rho$, column $(\gamma,\kappa)$ has exactly one $1$, at check $(\rho,\kappa+e_{\rho,\gamma})$; this gives (a) and (b). In~(c), $\gamma_1\ne\gamma_2$ by~(a). A column has one check per nonzero row block, so the two shared checks lie in distinct row blocks $\rho_1,\rho_2$, in which all four blocks are nonzero. Finally $\kappa_1+e_{\rho_i,\gamma_1}\equiv\kappa_2+e_{\rho_i,\gamma_2}$ for $i=1,2$, and the difference of these two congruences is the stated one.
\end{IEEEproof}

\begin{theorem}[QC-LDPC codes]\label{thm:qc}
Let $H$ be as above with every column of weight $w\ge2$, and let two columns $o_1,o_2$ share exactly $s\ge1$ checks, where $2s\le w$. Then for every $b$ with $w-2s\le b\le w-1$, problem~\eqref{eq:problem} with $t=1$ is infeasible, and
\[
\begin{aligned}
\degns\big(E^{\cnt}_{1,b}(H)\big)&\ge Z-w+s,\\
\degns\big(E^{\loc}_{1,b}(H)\big)&\le b+K+2 .
\end{aligned}
\]
In particular, if two columns share exactly one check (for instance if the girth is at least $6$ and some check has degree at least $2$), then $\degns(E^{\cnt}_{1,b}(H))\ge Z-w+1$ for $w-2\le b\le w-1$.
\end{theorem}
\begin{IEEEproof}
Every singleton has $\Phi=w>b$. Let $S^*=C(o_1)\cap C(o_2)$, and let $\mathcal J$ be the column block of $o_1$. By Lemma~\ref{lem:structure}(a), $\Phi(\{o_1,o_2\})=2w-2s$, so $\beta=w-2s\in[0,b]$. By Lemma~\ref{lem:qc}(a), the columns of $\mathcal J$ other than $o_1$ have pairwise disjoint check sets and avoid $C(o_1)$; and $o_2\notin\mathcal J$. By Lemma~\ref{lem:qc}(b), each of the $w-s$ checks of $C(o_2)\setminus S^*$ meets $\mathcal J$ in at most one column. So at least $M=Z-1-(w-s)$ columns of $\mathcal J$ also avoid $C(o_2)$, and they have weight $w$. If $M\ge1$, Corollary~\ref{cor:planted} gives $\degns\ge M+1=Z-w+s$; otherwise $Z-w+s\le1$ and the bound is trivial. The local bound is Theorem~\ref{thm:A} with $t=1$.
\end{IEEEproof}

The array codes of Fan~\cite{fan00} give an explicit family. For integers $w\ge2$, $L\ge2$ and $Z\ge2$, let $H(w,L,Z)$ be the $w\times L$ array of blocks $P^{\rho\gamma}$, $0\le\rho<w$, $0\le\gamma<L$; column $(\gamma,\kappa)$ meets the checks $(\rho,\kappa+\rho\gamma\bmod Z)$, $\rho<w$. For prime $Z$ and $L=Z$ this is the array code of~\cite{fan00}; for $L<Z$ its code is obtained by shortening.

\begin{theorem}[An explicit family]\label{thm:array}
Let $w\ge2$, $L\ge2$ and $Z>(w-1)(L-1)$, and let $H=H(w,L,Z)$. Then $H$ has $n=LZ$ columns of weight $w$ and $m=wZ$ checks of degree $L$, and no two columns share two checks. For $w-2\le b\le w-1$ and $K=\lceil\log_2(\lfloor L/2\rfloor+1)\rceil$,
\[
\begin{aligned}
Z-w+1\le\degns\big(E^{\cnt}_{1,b}(H)\big)&\le n+m+mK+b+4,\\
\degns\big(E^{\loc}_{1,b}(H)\big)&\le b+K+2 .
\end{aligned}
\]
For fixed $(w,L)$ and $Z\to\infty$, $\degns(E^{\cnt}_{1,b}(H))=\Theta(n)$ and $\degns(E^{\loc}_{1,b}(H))=O(1)$.
\end{theorem}
\begin{IEEEproof}
All blocks are nonzero, so every column has weight $w$ and, by Lemma~\ref{lem:qc}(b), every check meets each column block exactly once. If two columns shared two checks, Lemma~\ref{lem:qc}(c) would give $(\rho_1-\rho_2)(\gamma_1-\gamma_2)\equiv0\pmod Z$ with $0<|\rho_1-\rho_2|\le w-1$ and $0<|\gamma_1-\gamma_2|\le L-1$, which is impossible since $Z>(w-1)(L-1)$. The columns $o_1=(0,0)$ and $o_2=(1,0)$ have check sets $\{(\rho,0):\rho<w\}$ and $\{(\rho,\rho):\rho<w\}$, which share exactly $(0,0)$. Theorem~\ref{thm:qc} with $s=1$ gives the lower bound; explicitly, the planted columns are the $(0,\kappa)$ with $\kappa\notin\{0,\dots,w-1\}$, so $M=Z-w\ge0$. The upper bound is Proposition~\ref{prop:generic-upper} with $\sum_jK_j=mK$ and $t=1$, and the local bound is Theorem~\ref{thm:A}. Finally, $Z-w+1=n/L-w+1$, and since $m=wn/L$ and $b\le w-1$, the upper bound is at most $(1+w(1+K)/L)\,n+w+3$; both are linear in $n$ for fixed $(w,L)$.
\end{IEEEproof}

At $Z=(w-1)(L-1)$ the columns $(0,0)$ and $(L-1,0)$ share the two checks $(0,0)$ and $(w-1,0)$, so the threshold on $Z$ cannot be lowered uniformly in $(w,L)$.

\subsection{Planted $6$-Cycles and Larger Supports}
For column weight three a planted pair has odd $\beta$ by Lemma~\ref{lem:parity}, so $(t,b)=(1,0)$ is out of reach for single planted columns with target $Y+1$. Planted $6$-cycles have $\beta=0$.

\begin{theorem}[Planted $6$-cycles]\label{thm:triangle}
Let every column of $H$ have weight $3$, let the Tanner graph have girth at least $6$ and contain a $6$-cycle through the columns $o_1,o_2,o_3$, let the check degrees be at most $d$, and put $O=\{o_1,o_2,o_3\}$.
\begin{enumerate}
\item[(a)] For $t\in\{2,3\}$, problem~\eqref{eq:problem} is infeasible if $b\le2$ and feasible if $b\ge3$.
\item[(b)] For every $t\ge2$ and $b\ge0$, and any $M\ge1$ columns with pairwise disjoint check sets disjoint from $C(O)$, $\degns(E^{\cnt}_{t,b}(H))\ge M+1$. Such columns exist with $M\ge\lceil(n-3-6(d-1))/(1+3(d-1))\rceil$.
\item[(c)] For $H=H(3,L,Z)$ with $L\ge3$ and $Z>2(L-1)$, the columns $(1,0)$, $(0,2)$ and $(2,0)$ lie on a $6$-cycle, and $M=Z-4$ planted columns are available. So $\degns(E^{\cnt}_{t,b}(H))\ge Z-3$ for every $t\ge2$ and $b\ge0$, and, in the infeasible range $t\in\{2,3\}$, $b\le2$ of~(a),
\[
\degns\big(E^{\loc}_{t,b}(H)\big)\le t+b+K+1 .
\]
\end{enumerate}
\end{theorem}
\begin{IEEEproof}
(a) For $|S|\le3$, Lemma~\ref{lem:parity} gives $\Phi(S)=3|S|-2\sum_j\lfloor c_j(S)/2\rfloor$. Since $\lfloor c/2\rfloor\le\binom c2$ for every $c\ge0$, and two columns share at most one check,
\[
\textstyle\sum_j\lfloor c_j(S)/2\rfloor\le\sum_j\binom{c_j(S)}2\le\binom{|S|}2 .
\]
Hence $\Phi(S)\ge3$, $4$ and $3$ for $|S|=1$, $2$ and $3$, and the problem is infeasible for $b\le2$. A singleton has $\Phi=3$, so it is feasible for $b\ge3$.

(b) By Lemma~\ref{lem:structure}(b), $\Phi(O)=9-6=3$, so $\beta=0\le b$, and $|O|=3\le t+1$. Corollary~\ref{cor:planted} gives the bound, and Lemma~\ref{lem:structure}(c) with $|C(O)|=6$ and $n_3=n$ gives the count.

(c) In $H(3,L,Z)$ column $(\gamma,\kappa)$ meets row block $\rho$ at check $(\rho,\kappa+\rho\gamma)$. Thus $C((1,0))=\{(0,0),(1,1),(2,2)\}$, $C((0,2))=\{(0,2),(1,2),(2,2)\}$ and $C((2,0))=\{(0,0),(1,2),(2,4)\}$. Since $Z\ge5$, these are six distinct checks, and the three pairs share exactly $(2,2)$, $(1,2)$ and $(0,0)$, respectively, which closes a $6$-cycle. By Theorem~\ref{thm:array} the girth is at least $6$. Column $(0,\kappa)$ meets the checks $(0,\kappa)$, $(1,\kappa)$ and $(2,\kappa)$, so it avoids $C(O)$ exactly when $\kappa\notin\{0,1,2,4\}$. These $Z-4$ columns have pairwise disjoint check sets by Lemma~\ref{lem:qc}(a). Part~(b) gives the lower bound, and Theorem~\ref{thm:A} the local bound in the range where~(a) gives infeasibility.
\end{IEEEproof}

The local bound in~(c) cannot be extended to all $t\ge2$: once $t\ge\dmin$, problem~\eqref{eq:problem} with $b=0$ is feasible and both degrees are infinite. For example, $H(3,5,47)$ has $\dmin=6$ (Section~\ref{sec:md}), so for $t\ge6$ only the (then trivial) counting lower bound remains. For $4\le t<\dmin$ and $b=0$ the problem is infeasible and Theorem~\ref{thm:A} applies again; Section~\ref{sec:md} treats this range.

\subsection{Counting Versus Local Encodings}
Table~\ref{tab:compare} collects the comparisons proved in Sections~\ref{sec:sep}--\ref{sec:odd}. In every row problem~\eqref{eq:problem} is infeasible, and the counting lower bound is linear in the block length when the parameters in the first column are fixed; in the quasi-cyclic rows it is linear in $Z$, hence in $n$ for a fixed number of column blocks. For fixed $t$ and $b$ and bounded check degree, the local degree is bounded independently of the block length; the second cycle-code row, which allows $t$ to grow with $n$, instead has the explicit bound $t+b+2$. For small cycle codes the exact values are known: $\degns(E^{\cnt}_{1,0}(C_n))=3,4,4,5,5$ for $n=4,\dots,8$ (Table~\ref{tab:mixed}), against the planted bounds $1,2,2,3,3$.

\begin{table*}[t]
\caption{Counting versus local encodings: bounds proved in Sections~\ref{sec:sep}--\ref{sec:odd}}\label{tab:compare}
\centering\scriptsize\renewcommand{\arraystretch}{1.2}
\begin{tabular}{p{4.5cm}p{3.1cm}p{4.9cm}l}
\hline
Family and hypotheses & $(t,b)$ & $\degns\big(E^{\cnt}_{t,b}\big)$ & $\degns\big(E^{\loc}_{t,b}\big)$\\ \hline
$C_n$, $n\ge5$ & $(1,0)$ & $\ge\lfloor\frac{n-1}2\rfloor$, \ $\le3n+4$ & $=2$\\
$C_n$, $n\ge5$ & $1\le t\le n-1$, $b\le1$ & $\ge\lfloor\frac{n-1}2\rfloor$ & $\le t+b+2$\\
$(w,d)$-regular, girth $\ge6$ & $(1,b)$, $w-2\le b\le w-1$ & $\ge1+\big\lceil\frac{n-2-(2w-1)(d-1)}{1+w(d-1)}\big\rceil$ & $\le b+K+2$\\
$(w,d)$-regular, girth $\ge6$, an induced path or a star (e.g.\ girth $\ge8$) & $(1,b)$, $0\le b\le w-1$ & $\ge1+\big\lceil\frac{n/2-1-w^2(d-1)}{1+(2w-1)(d-1)}\big\rceil$ & $\le b+K+2$\\
CPM blocks, weight $w$, a pair sharing one check & $(1,b)$, $w-2\le b\le w-1$ & $\ge Z-w+1$ & $\le b+K+2$\\
$H(w,L,Z)$, $Z>(w-1)(L-1)$ & $(1,b)$, $w-2\le b\le w-1$ & $\ge Z-w+1$, \ $\le n+m+mK+b+4$ & $\le b+K+2$\\
$H(w,L,Z)$, $Z>(w-1)(L-1)$ & $(1,b)$, $0\le b\le w-1$ & $\ge1+\big\lceil\frac{LZ/2-1-w^2(L-1)}{1+(2w-1)(L-1)}\big\rceil$ & $\le b+K+2$\\
weight $3$, girth $\ge6$, a $6$-cycle & $t\in\{2,3\}$, $b\le2$ & $\ge1+\big\lceil\frac{n-3-6(d-1)}{1+3(d-1)}\big\rceil$ & $\le t+b+K+1$\\
$H(3,L,Z)$, $L\ge3$, $Z>2(L-1)$ & $t\in\{2,3\}$, $b\le2$ & $\ge Z-3$ & $\le t+b+K+1$\\
CPM array, weight $w$, minimum-weight codeword $\cw$ with $s$ check-disjoint columns & $\dmin-2s\le t<\dmin$, $b=0$ & $\ge Z-|C(\cw)|+1\ge Z-\lfloor w\dmin/2\rfloor+1$ & $\le t+K+1$\\
QC, codeword $\cw$ with difference set $D_\cw$ & $1\le t<\dmin$, $b=0$ & $\ge\lceil Z/|D_\cw|\rceil+1$ & $\le t+K+1$\\ \hline
\end{tabular}\\[2pt]
\parbox{\textwidth}{\scriptsize In every row problem~\eqref{eq:problem} is infeasible. A lower bound below $1$ is to be read as $1$. $K=\max_jK_j$ and $d$ is the maximum check degree. Rows 4 and 7: Corollaries~\ref{cor:E}--\ref{cor:odd}; rows 10 and 11: Theorem~\ref{thm:H} and Corollary~\ref{cor:D}, which are linear in $n$ when $\dmin$ is bounded, for example for $w\times L$ arrays with $L\ge w+1$ (Corollary~\ref{cor:md-asym}). By Proposition~\ref{prop:generic-upper}, $\degns(E^{\cnt}_{t,b})\le n+m+\sum_jK_j+t+b+3$ in every row, which is $O(n)$ whenever $m=O(n)$, the check degrees are bounded and $t$ and $b$ are fixed.}
\end{table*}

\section{Minimum-Distance Certification}\label{sec:md}
For $b=0$, infeasibility of problem~\eqref{eq:problem} means that $H$ has no nonzero codeword of weight at most $t$, that is, $\dmin\ge t+1$. Theorem~\ref{thm:B} gives two complementary ways of planting knapsacks from codewords: $O$ can be a codeword with a few columns removed and the planted sets single columns (Theorem~\ref{thm:H}), or $O$ can be empty and the planted sets whole codewords (Corollaries~\ref{cor:C} and~\ref{cor:D}). In this section $d$ denotes the weight of a codeword, not a check degree.

\subsection{Single Planted Columns from a Codeword}
\begin{theorem}[Minimum-distance planting]\label{thm:H}
Let $\cw\subseteq[n]$ be the support of a nonzero codeword of $H$, $d=|\cw|$. Let $S\subseteq\cw$ consist of $s\ge1$ columns with pairwise disjoint check sets, put $O=\cw\setminus S$ and $c=d-2s$, and assume $c\ge1$. Let $i_0,\dots,i_{M-1}$ be $M\ge1$ columns with pairwise disjoint check sets, all disjoint from $C(\cw)$, and let the columns of $S$ and the columns $i_v$ all have the same weight $w\ge1$.
\begin{enumerate}
\item[(a)] For every $t\ge c$ and $b\ge0$, $\degns\big(E^{\cnt}_{t,b}(H)\big)\ge M+1$. The bound is nontrivial when problem~\eqref{eq:problem} is infeasible; otherwise it holds by the convention $\degns=\infty$.
\item[(b)] If $\cw$ has minimum weight, $d=\dmin$, then problem~\eqref{eq:problem} with $b=0$ is infeasible exactly for $t\le\dmin-1$, so~(a) is a lower bound for certifying $\dmin\ge t+1$ for every $t$ with $\dmin-2s\le t\le\dmin-1$.
\item[(c)] If $H$ is a $w\times L$ array of $Z\times Z$ circulant permutation matrices, then the $i_v$ can be taken in any single column block, with
\[
M\ge Z-|C(\cw)|\ge Z-\Big\lfloor\frac{wd}{2}\Big\rfloor ,
\]
so $\degns\big(E^{\cnt}_{t,b}(H)\big)\ge Z-\lfloor wd/2\rfloor+1$ for every $t\ge c$ and $b\ge0$.
\end{enumerate}
\end{theorem}
\begin{IEEEproof}
(a) We apply Theorem~\ref{thm:B} with $O=\cw\setminus S$, $P_v=\{i_v\}$, $k=1$, $\beta=0$ and $c=d-2s\in\{1,\dots,t\}$. Every check has $c_j(\cw)$ even, and $c_j(S)\in\{0,1\}$ because no two columns of $S$ share a check. Hence $c_j(O)=c_j(\cw)-c_j(S)$ is odd exactly when $j\in C(S)$, so
\[
\Phi(O)=|C(S)|=sw .
\]
Moreover, every check of $C(\cw)$ has $c_j(\cw)\ge2$ and $c_j(S)\le1$, so it contains a column of $O$, and $C(O)=C(\cw)$. The columns $i_v$ avoid $C(\cw)$, so they lie outside $\cw$, and their check sets are pairwise disjoint, which is (R1); (R2) holds with $k=1$. Since $C(i_v)\cap C(O)=\emptyset$, adding $i_v$ to $O$ makes its $w$ checks odd, so $\lambda_v=w$ for every $v$, which is (R3). Finally,
\begin{align*}
k\,(\Phi(O)-\beta)&=sw,\\
\lambda\,(|O|-c)&=w\big[(d-s)-(d-2s)\big]=sw,
\end{align*}
which is (R4) with $c\in\{1,\dots,t\}$.

(b) If $d=\dmin$ and $b=0$, a support with $\Phi=0$ is a nonzero codeword, which has weight at least $\dmin$; so the problem is infeasible for $t\le\dmin-1$ and feasible for $t\ge\dmin$.

(c) Fix a column block $\mathcal J$. By Lemma~\ref{lem:qc}(a) its $Z$ columns have pairwise disjoint check sets, and by Lemma~\ref{lem:qc}(b) every check meets at most one of them, so at most $|C(\cw)|$ of them meet $C(\cw)$. Take the others as the $i_v$; they have weight $w$, and the bound is trivial if $Z\le|C(\cw)|$. Every check of $C(\cw)$ contains a positive even number of columns of $\cw$, and the columns of $\cw$ have $wd$ incidences in total, so $|C(\cw)|\le wd/2$.
\end{IEEEproof}

In (b), Theorem~\ref{thm:A} gives $\degns(E^{\loc}_{t,0}(H))\le t+K+1\le\dmin+K$ in the same range. For $s=1$, Theorem~\ref{thm:H} is Corollary~\ref{cor:planted}: $O$ is a codeword minus one column, $c=|O|-1$ and $r=-1$, and every codeword of weight $d\ge3$ admits $s=1$. For $s\ge2$, $r=-s$ and $|O|=d-s$ may exceed $t+1$; this is where the general target of Lemma~\ref{lem:A} is needed, and it extends the range of $t$ from $\{\dmin-2,\dmin-1\}$ down to $\dmin-2s$.

\subsection{The Packing Limit and the Choice of Codeword}
\begin{lemma}[Packing limit]\label{lem:packing}
In Theorem~\ref{thm:H}, $M\le\big\lfloor(m-|C(\cw)|)/w\big\rfloor$.
\end{lemma}
\begin{IEEEproof}
The $M$ planted columns have pairwise disjoint sets of $w$ checks, all outside $C(\cw)$.
\end{IEEEproof}
When the bound is attained, the planted columns cover every check outside $C(\cw)$, and no larger planting of this kind exists for that codeword. This is an optimality statement about the planting configuration, not about the Nullstellensatz degree: the true degree may be larger than $M+1$ (Section~\ref{sec:limits}).

The capacity depends on the codeword, not only on its weight: codewords of the same weight can differ in $|C(\cw)|$, since a check may contain four or more of their columns, and in how $C(\cw)$ is distributed over the row blocks, which constrains the packing in each column block. Example~\ref{ex:B} has two classes of minimum-weight codewords with $|C(\cw)|=11$ and $12$ and optimal plantings of $18$ and $17$ columns, so the codeword should be chosen only after its weight, neighbourhood and packing have been verified.

\subsection{An Asymptotic Separation for Arrays of Circulants}
\begin{corollary}[Certifying the minimum distance]\label{cor:md-asym}
Fix $w\ge2$ and $L\ge w+1$. For $Z\ge2$ let $H_Z$ be a $w\times L$ array of $Z\times Z$ circulant permutation matrices whose Tanner graph has girth at least $6$, with $n=LZ$, $m=wZ$, check degree $L$ and $K=\lceil\log_2(\lfloor L/2\rfloor+1)\rceil$. Let $d_Z=\dmin(H_Z)$. Then $w+1\le d_Z\le(w+1)!$~\cite[Cor.~9]{sv12}, and
\begin{align*}
\degns\big(E^{\cnt}_{d_Z-1,0}(H_Z)\big)&\ge Z-\big\lfloor w\,(w+1)!/2\big\rfloor+1,\\
\degns\big(E^{\cnt}_{d_Z-1,0}(H_Z)\big)&\le n+m+mK+d_Z+2,\\
\degns\big(E^{\loc}_{d_Z-1,0}(H_Z)\big)&\le d_Z+K\le(w+1)!+K.
\end{align*}
Hence, for fixed $(w,L)$, certifying the minimum distance needs counting degree $\Theta(n)$, while the local degree is $O(1)$. This applies to $H(w,L,Z)$ for every $Z>(w-1)(L-1)$.
\end{corollary}
\begin{IEEEproof}
Let $\cw$ be a nonzero codeword and $i\in\cw$. Each of the $w$ checks of $i$ contains a second column of $\cw$, and these columns are distinct because two columns share at most one check; hence $d_Z\ge w+1\ge3$. By~\cite[Cor.~9]{sv12}, a $w\times L$ array of circulant permutation matrices with $L\ge w+1$ has minimum distance at most $(w+1)!$: it is a type-1 quasi-cyclic code whose weight matrix is the all-one $w\times L$ matrix, and in the bound of~\cite[Th.~8]{sv12}, a minimum over sets of $w+1$ column blocks (which requires $L\ge w+1$), each of the $w+1$ permanents of $w\times w$ all-one matrices equals $w!$. The bound goes back to~\cite{mackay01}; see also~\cite[Sec.~V]{fossorier04}. Take a minimum-weight codeword and $s=1$, so $c=d_Z-2\ge1$ and $t=d_Z-1\ge c$; Theorem~\ref{thm:H}(c) gives the lower bound, since $\lfloor wd_Z/2\rfloor\le\lfloor w(w+1)!/2\rfloor$. The upper bound is Proposition~\ref{prop:generic-upper} with $t=d_Z-1$, $b=0$ and $\sum_jK_j=mK$, and the local bound is Theorem~\ref{thm:A}. The parameters $t=d_Z-1\le(w+1)!-1$, $b=0$ and the check degree $L$ are bounded, so both statements are uniform in $Z$. For $H(w,L,Z)$ the girth condition holds by Theorem~\ref{thm:array}.
\end{IEEEproof}

The corollary uses only $s=1$; larger $s$ give the bound for more values of $t$ but not a larger leading term. Without a bound on $\dmin$ that is uniform in $Z$, Theorem~\ref{thm:H}(c) still holds for each code, but it need not be linear in $n$; we make no asymptotic claim for $L\le w$.

\subsection{Planted Codewords}
\begin{corollary}[Planted codewords]\label{cor:C}
Let $P_0,\dots,P_{M-1}$ be pairwise disjoint supports of nonzero codewords of a common weight $k$, such that no check meets two of them. Then $\degns(E^{\cnt}_{t,b}(H))\ge M+1$ for every $t\ge1$ and $b\ge0$. In particular this holds, with problem~\eqref{eq:problem} infeasible, for $b=0$ and every $t<\dmin$.
\end{corollary}
\begin{IEEEproof}
Take $O=\emptyset$, $\beta=0$ and $c=1$. A codeword support has $\Phi(P_v)=0$, so $\lambda=0$, and (R4) reads $0=0$. (R1) and (R2) hold by assumption. Here $r=1/k$.
\end{IEEEproof}
The substitution exposes a knapsack with an unattainable target: the support size is a multiple of $k$, while the counter says that it equals $c=1$. Corollary~\ref{cor:C} needs no structure inside the codewords and serves every $(t,b)$ at once, including $t=1$, where Theorem~\ref{thm:H} does not apply for codewords of even weight.

\begin{corollary}[Shifted codewords in QC codes]\label{cor:D}
Let $H$ be an array of $Z\times Z$ blocks, each zero or a circulant permutation matrix, without zero columns, and let $\cw$ be the support of a nonzero codeword. Let
\[
D_\cw=\{a-a'\bmod Z:\ (\rho,a),(\rho,a')\in C(\cw)\text{ for some }\rho\},
\]
and let $n_\rho$ be the number of checks of $C(\cw)$ in row block $\rho$. Then the cyclic shifts of $\cw$ contain at least $\lceil Z/|D_\cw|\rceil$ codewords no check meets two of, and $|D_\cw|\le\delta:=1+\sum_\rho n_\rho(n_\rho-1)$. Hence $\degns(E^{\cnt}_{t,b}(H))\ge\lceil Z/|D_\cw|\rceil+1$ whenever problem~\eqref{eq:problem} is infeasible.
\end{corollary}
\begin{IEEEproof}
Column $(\gamma,\kappa)$ meets check $(\rho,\kappa+e_{\rho,\gamma})$ in every nonzero block, so the shift $\kappa\mapsto\kappa+\sigma$ in every column block maps codewords to codewords and moves every check $(\rho,a)$ to $(\rho,a+\sigma)$. The shifts of $\cw$ by $\sigma$ and $\sigma'$ share a check if and only if $\sigma-\sigma'\in D_\cw$. Choosing shifts greedily in $\mathbb Z_Z$, each choice excludes at most $|D_\cw|$ residues, which gives the count; check-disjoint shifts are also disjoint as sets, since $H$ has no zero column. $D_\cw$ contains $0$ and at most $n_\rho(n_\rho-1)$ further differences from row block $\rho$. Corollary~\ref{cor:C} completes the proof.
\end{IEEEproof}

\subsection{Examples}
For prime $Z\ge5$ and $L=Z$, $H(3,L,Z)$ is a parity-check matrix of the full array code of column weight three~\cite{fan00}, whose minimum distance is $6$~\cite{yang03} (see also~\cite{rosnes14}, which summarizes the known values); an exhaustive search confirms $\dmin=6$ for every prime $Z$ from $5$ to $23$. For $L<Z$ the code is a shortening of the full code, so its minimum distance is at least $6$. None of this is used below: the minimum distance is computed exactly in each example. All codewords, packings and bounds below were computed and verified exactly (Section~\ref{sec:verif}); the configurations are in the supplementary material, and Table~\ref{tab:md} collects them. Columns are written as $(\gamma,\kappa)$.

\begin{example}[$H(3,5,47)$]\label{ex:A}
Here $n=235$, $m=141$ and $K=2$. Exhaustive search gives $\dmin=6$; the $47$ minimum-weight codewords are the cyclic shifts of
\[
\cw=\{(0,0),(1,0),(1,45),(3,43),(3,45),(4,43)\},
\]
and each has $|C(\cw)|=9$, with every check of $C(\cw)$ containing two columns of $\cw$. With $S=\{(0,0),(3,43)\}$, which share no check, $s=2$ and $c=2$, CP-SAT finds $44$ planted columns and proves that no larger family exists. Since
\[
44=\frac{141-9}{3},
\]
the planting is perfect: the planted columns cover every check outside $C(\cw)$, and Lemma~\ref{lem:packing} shows its optimality. Theorem~\ref{thm:H} gives
\[
\degns\big(E^{\cnt}_{t,0}(H(3,5,47))\big)\ge45\qquad(2\le t\le5),
\]
while Theorem~\ref{thm:A} gives $\degns(E^{\loc}_{t,0})\le t+3$, which is $5$ at $t=2$ and $8$ at $t=5$. Theorem~\ref{thm:H}(c) guarantees at least $47-9=38$ planted columns in every single column block, hence the bound $39$; in fact column blocks $1$, $2$ and $3$ each contain $42$ columns avoiding $C(\cw)$, which gives $43$ from one block. Planted codewords give less here: $15$ pairwise check-disjoint minimum-weight codewords exist and no more (CP-SAT, and $\lfloor141/9\rfloor=15$), so Corollary~\ref{cor:C} gives $16$, for every $t\le5$ including $t=1$. On $H(3,5,Z)$, $|D_\cw|=7$ and $\delta=19$, so Corollary~\ref{cor:D} guarantees $\lceil Z/7\rceil$ shifts; greedy choices give $\lfloor Z/3\rfloor$, that is, $3$, $7$ and $15$ for $Z=11,23,47$, which is optimal because each shift occupies three checks of row block $0$.
\end{example}

\begin{example}[$H(3,4,23)$]\label{ex:B}
Here $n=92$, $m=69$ and $K=2$, and exhaustive search gives $\dmin=8$. There are $46$ minimum-weight codewords, the shifts of two codewords; $23$ of them have $|C(\cw)|=11$ and $23$ have $|C(\cw)|=12$. The codeword
\[
\cw=\{(0,3),(1,1),(1,2),(1,3),(2,0),(2,1),(2,2),(3,0)\}
\]
has weight $8$ and $|C(\cw)|=11$: the check $(1,3)$ contains the four columns $(0,3)$, $(1,2)$, $(2,1)$ and $(3,0)$, and the other ten checks contain two columns each. With $S=\{(0,3),(2,0),(2,2)\}$, $s=3$ and $c=2$, CP-SAT finds an optimal planting of $18$ columns, one less than the capacity $\lfloor(69-11)/3\rfloor=19$. Hence
\[
\degns\big(E^{\cnt}_{t,0}(H(3,4,23))\big)\ge19\qquad(2\le t\le7),
\]
while $\degns(E^{\loc}_{t,0})\le t+3$, which is $10$ at $t=7$. Theorem~\ref{thm:H}(c) guarantees $23-11=12$ planted columns in every single column block, hence the bound $13$; in fact column blocks $1$ and $2$ each contain $18$ columns avoiding $C(\cw)$, so one block already attains the optimum. A codeword with $|C(\cw)|=12$ admits at most $17$ planted columns (optimal), so the choice of codeword matters. Corollary~\ref{cor:C} gives $6$ from $5$ pairwise check-disjoint minimum-weight codewords (optimal).
\end{example}

\begin{table*}[t]
\caption{Minimum-distance and odd-weight examples ($K=2$ in all three codes)}\label{tab:md}
\centering\scriptsize\renewcommand{\arraystretch}{1.12}\setlength{\tabcolsep}{4.0pt}
\begin{tabular}{llcccccc}
\hline
Code & Construction & $(t,b)$ & $M$ & status & capacity & $\degns(E^{\cnt}_{t,b})$ & $\degns(E^{\loc}_{t,b})$\\ \hline
$H(3,5,47)$, & Theorem~\ref{thm:H}, $s=2$, $|C(\cw)|=9$ & $2\le t\le5$, $b=0$ & $44$ & optimal, perfect & $44$ & $\ge45$ & $\le t+3$\\
$\dmin=6$ & Theorem~\ref{thm:H}(c), one column block & $2\le t\le5$, $b=0$ & $\ge38$ & --- & --- & $\ge39$ & $\le t+3$\\
 & Corollary~\ref{cor:C}, planted codewords & $1\le t\le5$, $b=0$ & $15$ & optimal & $15$ & $\ge16$ & $\le t+3$\\
$H(3,4,23)$, & Theorem~\ref{thm:H}, $s=3$, $|C(\cw)|=11$ & $2\le t\le7$, $b=0$ & $18$ & optimal & $19$ & $\ge19$ & $\le t+3$\\
$\dmin=8$ & Theorem~\ref{thm:H}, $s=3$, $|C(\cw)|=12$ & $2\le t\le7$, $b=0$ & $17$ & optimal & $19$ & $\ge18$ & $\le t+3$\\
 & Theorem~\ref{thm:H}(c), one column block & $2\le t\le7$, $b=0$ & $\ge12$ & --- & --- & $\ge13$ & $\le t+3$\\
 & Corollary~\ref{cor:C}, planted codewords & $1\le t\le7$, $b=0$ & $5$ & optimal & $6$ & $\ge6$ & $\le t+3$\\
 & Corollary~\ref{cor:E}, star & $(1,b)$, $b\le2$ & $12$ & optimal, perfect & $12$ & $\ge13$ & $\le b+4$\\
 & Corollary~\ref{cor:E}, induced path & $(1,b)$, $b\le2$ & $11$ & optimal & $12$ & $\ge12$ & $\le b+4$\\
$H(4,5,31)$ & Corollary~\ref{cor:E}, star & $(1,b)$, $b\le3$ & $14$ & optimal & $15$ & $\ge15$ & $\le b+4$\\
 & Corollary~\ref{cor:E}, induced path & $(1,b)$, $b\le3$ & $14$ & optimal & $15$ & $\ge15$ & $\le b+4$\\ \hline
\end{tabular}\\[2pt]
\parbox{\textwidth}{\scriptsize $M$: number of planted sets; for Theorem~\ref{thm:H}(c), the guaranteed number $Z-|C(\cw)|$ in any column block. Status: ``optimal'' means that CP-SAT proved that no larger family of the same kind exists for the chosen $\cw$ or $O$; ``perfect'' means that the planted sets and $C(\cw)$ (or $C(O)$) cover all checks. Capacity: $\lfloor(m-|C(\cw)|)/w\rfloor$ for Theorem~\ref{thm:H}, $\lfloor m/\min|C(\cw)|\rfloor$ for Corollary~\ref{cor:C}, and $\lfloor(m-w^2)/(2w-1)\rfloor$ for Corollary~\ref{cor:E}. Every counting bound is $M+1$ and was verified generator by generator on the full counting encoding; every local bound is Theorem~\ref{thm:A}. All problems in the table are infeasible.}
\end{table*}

\section{Odd Column Weight: Planted Pairs, Paths and Stars}\label{sec:odd}
By Lemmas~\ref{lem:parity} and~\ref{lem:G}, planted sets of odd size cannot reach $(t,b)=(1,0)$ when the column weight is odd. Planted pairs can.

\subsection{Planted Pairs}
A \emph{domino} is a pair of columns of weight $w$ that share exactly one check; it has $\Phi=2w-2$ and $|C(\cdot)|=2w-1$.

\begin{corollary}[Dominoes]\label{cor:E}
Let $O\subseteq[n]$ satisfy $|O|=w+1$ and $\Phi(O)=w(w-1)$, and let $P_0,\dots,P_{M-1}$ be dominoes of weight-$w$ columns whose check sets $C(P_v)$ are pairwise disjoint and disjoint from $C(O)$. Then $\degns(E^{\cnt}_{t,b}(H))\ge M+1$ for every $t\ge1$ and $b\ge0$. If every column has weight $w\ge2$, problem~\eqref{eq:problem} with $t=1$ is infeasible for $0\le b\le w-1$, and then $\degns(E^{\loc}_{1,b}(H))\le b+K+2$.
\end{corollary}
\begin{IEEEproof}
Since $C(P_v)\cap C(O)=\emptyset$, $\lambda_v=\Phi(P_v)=2w-2$ for every $v$. Take $k=2$, $c=1$ and $\beta=0$; then (R4) reads $2w(w-1)=(2w-2)\,w$, and $r=(1-(w+1))/2=-w/2$. (R1) and (R2) hold by assumption, and Theorem~\ref{thm:B} gives the lower bound. Every singleton has $\Phi=w>b$, and the local bound is Theorem~\ref{thm:A} with $t=1$.
\end{IEEEproof}

\begin{lemma}[Paths and stars]\label{lem:pathstar}
Let every column have weight $w\ge2$. Each of the following sets $O$ has $|O|=w+1$, $|C(O)|=w^2$ and $\Phi(O)=w(w-1)$.
\begin{enumerate}
\item[(a)] An \emph{induced path} $O=\{o_0,\dots,o_w\}$: consecutive columns share exactly one check and nonconsecutive columns share none.
\item[(b)] A \emph{star} $O=\{o,\ell_1,\dots,\ell_w\}$: if $C(o)=\{j_1,\dots,j_w\}$, then $\ell_i$ shares with $o$ only the check $j_i$, and no two leaves share a check.
\end{enumerate}
\end{lemma}
\begin{IEEEproof}
(a) The check shared by $o_{i-1}$ and $o_i$ contains no other column of $O$, since such a column would share it with a nonconsecutive one; so the $w$ shared checks are distinct and have $c_j(O)=2$, and every other check of $C(O)$ has $c_j(O)=1$. Since $\sum_jc_j(O)=w(w+1)$, $|C(O)|=w(w+1)-w=w^2$ and $\Phi(O)=w^2-w$. (b) The checks $j_i$ contain $o$ and $\ell_i$ and no other leaf, so $c_{j_i}(O)=2$; the remaining $w(w-1)$ checks of the leaves are distinct and have $c_j(O)=1$. Hence $|C(O)|=w+w(w-1)=w^2$ and $\Phi(O)=w(w-1)$.
\end{IEEEproof}

\subsection{When Paths and Stars Exist}
\begin{proposition}[Existence]\label{prop:exist}
Let every column have weight $w\ge2$.
\begin{enumerate}
\item[(a)] If the Tanner graph has girth at least $8$ and every check has degree at least $2$, every column is the centre of a star.
\item[(b)] In $H(w,L,Z)$ with $L\ge2$ and $Z\ge w$, the column $(0,0)$ and the leaves $\ell_\rho=(1,-\rho\bmod Z)$, $0\le\rho<w$, form a star.
\item[(c)] In the point--line incidence matrix of a finite projective plane~\cite{kou01}, whose Tanner graph has girth $6$, every two columns share exactly one check. So there is no induced path on three columns and no star, and no two dominoes have disjoint check sets.
\end{enumerate}
\end{proposition}
\begin{IEEEproof}
(a) Let $C(o)=\{j_1,\dots,j_w\}$ and pick $\ell_i\in N(j_i)\setminus\{o\}$. Girth at least $6$ implies that $\ell_i$ shares no other check with $o$ and that the $\ell_i$ are distinct. If $\ell_i$ and $\ell_{i'}$ shared a check $j$, then $o,j_i,\ell_i,j,\ell_{i'},j_{i'}$ would be a $6$-cycle. (b) Column $(\gamma,\kappa)$ meets the checks $(\rho',\kappa+\rho'\gamma)$, so $C((0,0))=\{(\rho',0)\}$ and $C(\ell_\rho)=\{(\rho',\rho'-\rho)\}$. For $0\le\rho,\rho'<w\le Z$, $\rho'-\rho\equiv0$ only if $\rho'=\rho$, so $\ell_\rho$ shares exactly the check $(\rho,0)$ with $(0,0)$. The leaves are distinct columns of column block $1$, so their check sets are disjoint by Lemma~\ref{lem:qc}(a). (c) Two distinct lines of a projective plane meet in exactly one point, so every two columns share one check. In an induced path $o_0$ and $o_2$ would share a check, and in a star two leaves would; and columns $a\in P_0$, $a'\in P_1$ of two dominoes share a check.
\end{IEEEproof}

In particular, girth six alone guarantees neither structure, and the projective-plane codes, for which every two columns are adjacent, are a counterexample to such a claim. Whether their counting degree grows at $(t,b)=(1,0)$ is open (Section~\ref{sec:disc}).

\subsection{Counting Dominoes}
\begin{corollary}[Linear lower bounds for every column weight]\label{cor:odd}
Let $H$ be $(w,d)$-regular with $w,d\ge2$, let its Tanner graph have girth at least $6$, and let $H$ contain an induced path or a star $O$. Then there are at least
\[
M\ge\Big\lceil\frac{n/2-1-w^2(d-1)}{1+(2w-1)(d-1)}\Big\rceil
\]
dominoes as in Corollary~\ref{cor:E}, so for $0\le b\le w-1$
\[
\begin{aligned}
\degns\big(E^{\cnt}_{1,b}(H)\big)&\ge1+\max\{0,M\},\\
\degns\big(E^{\loc}_{1,b}(H)\big)&\le b+K+2 .
\end{aligned}
\]
This applies to every $(w,d)$-regular matrix of girth at least $8$, and to $H(w,L,Z)$ with $L\ge2$ and $Z>(w-1)(L-1)$, where $d=L$ and $n=LZ$. For fixed $(w,d)$ the counting degree is $\Theta(n)$ and the local degree $O(1)$, for every column weight, including odd $w$ with $b=0$.
\end{corollary}
\begin{IEEEproof}
By the girth condition every pair of columns in a common check is a domino, so there are $nw(d-1)/2$ dominoes. The $2w-1$ checks of a domino contain at most $1+(2w-1)(d-1)$ columns, and each column lies in $w(d-1)$ dominoes, so a domino is incompatible with at most $\Delta=w(d-1)\big(1+(2w-1)(d-1)\big)$ dominoes, itself included. Each check of $C(O)$ contains $c_j(O)\ge1$ columns of $O$, so the $w^2$ checks of $C(O)$ contain at most $|O|+\sum_{j\in C(O)}(d-c_j(O))=1+w^2(d-1)$ columns, and at most $w(d-1)\big(1+w^2(d-1)\big)$ dominoes meet $C(O)$. Choosing dominoes greedily gives $M\ge\lceil(nw(d-1)/2-w(d-1)(1+w^2(d-1)))/\Delta\rceil$, which is the stated bound. Corollary~\ref{cor:E} gives the degree bounds. For girth at least $8$, Proposition~\ref{prop:exist}(a) provides a star; for $H(w,L,Z)$, Theorem~\ref{thm:array} gives girth at least $6$ and Proposition~\ref{prop:exist}(b) a star, since $Z>(w-1)(L-1)\ge w-1$. The upper bound $O(n)$ is Proposition~\ref{prop:generic-upper}.
\end{IEEEproof}

So the counting degree at $(t,b)=(1,0)$ grows linearly for every odd column weight on the array codes $H(w,L,Z)$ and on regular families of girth at least eight, which single planted columns cannot reach (Lemma~\ref{lem:parity}); Corollary~\ref{cor:odd} says nothing about girth-six families without induced paths or stars. For arrays with bounded minimum distance, such as $w\times L$ arrays with $L\ge w+1$, Corollaries~\ref{cor:C} and~\ref{cor:D} give a second linear bound at $(t,b)=(1,0)$. The same dominoes give $\degns(E^{\cnt}_{t,b})\ge M+1$ for every $t\ge1$, and with Theorem~\ref{thm:triangle}(a) they cover $t\in\{2,3\}$, $b\le2$ for column weight three without planted $6$-cycles.

\begin{example}[Stars and paths on array codes]\label{ex:C}
On $H(3,4,23)$ the star with centre $(0,0)$ and leaves $(1,0)$, $(1,22)$ and $(1,21)$ has $|C(O)|=9$ and $\Phi(O)=6$. CP-SAT finds $12$ dominoes compatible with it and with each other, and $12$ is optimal: together with $C(O)$ they cover $9+12\cdot5=69=m$ checks, a perfect packing. Hence
\[
\degns\big(E^{\cnt}_{1,b}(H(3,4,23))\big)\ge13\qquad(0\le b\le2),
\]
against $\degns(E^{\loc}_{1,b})\le b+4$. The induced path $(0,0),(1,0),(0,1),(1,1)$ also has $|C(O)|=9$, but it admits only $11$ dominoes (optimal), which gives $12$: the path and the star block different checks. On $H(4,5,31)$ the star with centre $(0,0)$ and leaves $(1,0),(1,30),(1,29),(1,28)$ and the induced path $(0,0),(1,0),(0,1),(1,1),(0,4)$ both admit $14$ dominoes (optimal, against the capacity $\lfloor(124-16)/7\rfloor=15$), so
\[
\degns\big(E^{\cnt}_{1,b}(H(4,5,31))\big)\ge15\qquad(0\le b\le3),
\]
against $\degns(E^{\loc}_{1,b})\le b+4$. The greedy count of Corollary~\ref{cor:odd} guarantees only $2$ and $1$ dominoes on these two codes; it is linear in $n$ but weak for small $Z$.
\end{example}

\section{Mixed Encodings and the Two-Counter Phenomenon}\label{sec:mixed}

\subsection{Four Exact Encodings}
Each of the two cardinality bounds in~\eqref{eq:feas} can be imposed locally (L) or by counting (C). For $\mathrm{XY}\in\{\mathrm{LL},\mathrm{LC},\mathrm{CL},\mathrm{CC}\}$, where the first letter refers to the support bound and the second to the syndrome bound, $\Emix{XY}{t,b}(H)$ contains $x$, $u$, the bits $q_{j,k}$, their Boolean axioms and the generators $L_j$, together with:
\begin{itemize}
\item for a local support bound, $P_t$ and the obstructions $x_T$, $|T|=t+1$; for a counted one, $y_0,\dots,y_t$ with their Boolean axioms, $W$, and the first line of~\eqref{eq:cnt};
\item for a local syndrome bound, the obstructions $u_R$, $|R|=b+1$; for a counted one, $z_0,\dots,z_b$ with their Boolean axioms, $U$, and the second line of~\eqref{eq:cnt}.
\end{itemize}
Thus $\Emix{LL}{t,b}=E^{\loc}_{t,b}$ and $\Emix{CC}{t,b}=E^{\cnt}_{t,b}$, while $\Emix{CL}{t,b}$ counts the support and keeps the syndrome local, and $\Emix{LC}{t,b}$ does the opposite. By the argument of Proposition~\ref{prop:exact}, all four are exact encodings of~\eqref{eq:problem}.

\subsection{One Counter Gives Constant Degree}
For $(t,b)=(1,0)$ on $C_n$, $\Emix{CL}{1,0}(C_n)$ contains $W-\sum_ix_i$, $y_0+y_1-1$, $W-y_1$, $y_0$ and the obstructions $u_j$, $j\in\Zn$; $\Emix{LC}{1,0}(C_n)$ contains $P_1$, the obstructions $x_ix_k$, and $U-\sum_ju_j$, $z_0-1$, $U$. Note that $\sum_ju_j=U-(U-\sum_ju_j)\in I_1(\Emix{LC}{1,0})$ and
\[
{\textstyle\sum_ix_i}-1=(W-y_1)-\big(W-{\textstyle\sum_ix_i}\big)+(y_0+y_1-1)-y_0\in I_1\big(\Emix{CL}{1,0}\big).
\]

\begin{proposition}[One counter is not enough]\label{prop:D}
For every $n\ge3$,
\[
\degns\big(\Emix{CL}{1,0}(C_n)\big)\le4,\qquad \degns\big(\Emix{LC}{1,0}(C_n)\big)\le3 .
\]
\end{proposition}
\begin{IEEEproof}
Write $\tilde L_j=L_j+u_j=x_{j-1}+x_j-2q_j$.

\emph{$\Emix{CL}{1,0}$.} Here $\tilde L_j\in I_1$, and the identity
\[
x_{j-1}+x_j-2x_{j-1}x_j=\Bool{x_{j-1}}+\Bool{x_j}-4\Bool{q_j}-(x_{j-1}+x_j+2q_j-2)\tilde L_j
\]
gives $x_{j-1}+x_j-2x_{j-1}x_j\in I_2$. Then
\[
x_{j-1}-x_{j-1}x_j=x_{j-1}(x_{j-1}+x_j-2x_{j-1}x_j)-\Bool{x_{j-1}}(1-2x_j)\in I_3,
\]
and in the same way $x_j-x_{j-1}x_j\in I_3$; hence $x_{j-1}-x_j\in I_3$ for every $j$. Telescoping gives $x_i-x_0\in I_3$ for all $i$, so $nx_0-1=(\sum_ix_i-1)-\sum_i(x_i-x_0)\in I_3$. Multiplying by $x_0$ gives $x_0(nx_0-1)=n\Bool{x_0}+(n-1)x_0\in I_4$, so $x_0\in I_4$, and $1=nx_0-(nx_0-1)\in I_4$.

\emph{$\Emix{LC}{1,0}$.} Here $x_{j-1}x_j$ is an obstruction. With $s_j=x_{j-1}+x_j$,
\[
2q_j(1+2u_j)=\Bool{x_{j-1}}+\Bool{x_j}+2x_{j-1}x_j-L_j(2s_j-L_j-1)-\Bool{u_j}-4\Bool{q_j}\in I_2 ,
\]
and $q_j=q_j(1+2u_j)(1-\tfrac23u_j)+\tfrac43\,q_j\Bool{u_j}$ gives $q_j\in I_3$. Then $u_j=s_j-2q_j-L_j\in s_j+I_3$, so $2\sum_ix_i=\sum_js_j\in\sum_ju_j+I_3=I_3$. Hence $1=P_1+\sum_ix_i\in I_3$.
\end{IEEEproof}

\begin{table}[t]
\caption{Nullstellensatz degrees for $(t,b)=(1,0)$ on $C_n$}\label{tab:mixed}
\centering\scriptsize\setlength{\tabcolsep}{3.1pt}
\begin{tabular}{lccccccc}
\hline
$n$ & 4 & 5 & 6 & 7 & 8 & 10 & 12\\ \hline
$\OPP_n$ & 3 & 4 & 4 & 5 & 5 & 6 & 7\\
$\Emix{CC}{1,0}$ (count both) & 3 & 4 & 4 & 5 & 5 & $\ge5$ & $\ge5$\\
$\Emix{CL}{1,0}$ (count support) & 3 & 3 & 4 & 4 & 4 & 4 & 4\\
$\Emix{LC}{1,0}$ (count syndrome) & 3 & 3 & 3 & 3 & 3 & 3 & 3\\
$\Emix{LL}{1,0}$ (count neither) & 2 & 2 & 2 & 2 & 2 & 2 & 2\\ \hline
\end{tabular}\\[2pt]
\parbox{\columnwidth}{\scriptsize All entries were computed exactly over $\Q$ (Section~\ref{sec:verif}); an entry $\ge v$ is a lower bound, every other entry is the exact degree. Lower bounds come from rational functionals as in Lemma~\ref{lem:dual}; upper bounds from explicit certificates, from Proposition~\ref{prop:D} for $\Emix{LC}{}$ and $\Emix{CL}{}$, and from Corollary~\ref{cor:B} for $\Emix{LL}{}$. For $n=10,12$ the lower bounds agree with~\cite{itw27}.}
\end{table}

\subsection{Why the Growth Needs Both Counters}
When the syndrome bound is local, every $u_j$ is a generator, the relations $x_{j-1}=x_j$ propagate along the cycle at constant degree, and the single constraint $\sum_ix_i=1$ is easy to contradict. When the support bound is local, the pairwise obstructions make $q_j$, and hence $u_j-s_j$, cheap, and the syndrome count reduces to the support count. Only when both bounds are counted is the remaining global information a pair of sums, $\sum_ix_i=1$ and $\sum_ju_j=0$, into which a knapsack can be planted. The planted substitutions send some local obstruction to a nonzero element (for example $u_j\mapsto\eta_v$ for $j\in C(i_v)$, or $x_{o_1}x_{o_2}\mapsto1$), so they cannot be applied when either bound is local. Whether the mixed encodings have bounded degree beyond $C_n$ is open.

\section{Optimality and Limits of the Planting Method}\label{sec:limits}
Theorem~\ref{thm:B} gives degree $M+1$ for $M$ planted sets. This section shows that this is the best any argument of the same kind can give, that $M+1$ is attained, and what the method cannot see.

\subsection{Limits of the Knapsack Argument}
\begin{corollary}[Limits of the knapsack argument]\label{cor:knapsack-limit}
Under the hypotheses of Theorem~\ref{thm:B}, with $r\notin\{0,\dots,M\}$:
\begin{enumerate}
\item[(a)] $\Lambda_{M,r}\circ\tau$ vanishes on every product $\mu g$ of degree at most $M$ and not on every product of degree $M+1$, by~\eqref{eq:nonvac-planted};
\item[(b)] no linear functional $\tilde\Lambda$ on $\Q[\eta]$ with $\tilde\Lambda(1)=1$ makes $\tilde\Lambda\circ\tau$ vanish on all products $\mu g$ of degree at most $M+1$;
\item[(c)] these are limits of the method, not of the degree: $\degns(E^{\cnt}_{1,0}(C_5))=4$ (Table~\ref{tab:mixed}), while every planting found on $C_5$ has $M=1$ (Corollary~\ref{cor:lin-cycle}(b) and Section~S-III).
\end{enumerate}
\end{corollary}
\begin{IEEEproof}
(b) Fix $i_v\in P_v$. Since $\tau(x_{i_v})=\eta_v$, every monomial $\eta^\alpha$ is the image of a monomial of the same degree in the $x_{i_v}$, and $\Bool{\eta_v}=\tau(\Bool{x_{i_v}})$; moreover $\tau(W-\sum_ix_i)=-kT$ with $k\ne0$. A functional as in~(b) would therefore annihilate $\eta^\alpha\Bool{\eta_v}$ for $|\alpha|\le M-1$ and $\eta^\alpha T$ for $|\alpha|\le M$, and applying it to the certificate of Proposition~\ref{prop:target}(a) would give $1=0$.
\end{IEEEproof}

\subsection{A Cap for Pullback Arguments}
\begin{theorem}[A cap for pullback arguments]\label{thm:cap}
Let $E$ be a system over $\Q$ without common zero whose generators other than Boolean axioms have degree at most $d_G$. Let $\tau$ be a ring homomorphism from the polynomial ring of $E$ to $\Q[\zeta_1,\dots,\zeta_N]$ such that
\begin{enumerate}
\item[(C1)] every $\zeta_k$ equals $\tau(v)$ for some Boolean variable $v$ of $E$;
\item[(C2)] $\tau(\Bool v)\in\langle\Bool{\zeta_1},\dots,\Bool{\zeta_N}\rangle$ for every Boolean variable $v$ of $E$;
\item[(C3)] $\deg\tau(g)\le d_G$ for every generator $g$ that is not a Boolean axiom.
\end{enumerate}
If a linear functional $\tilde\Lambda$ on $\Q[\zeta]$ satisfies $\tilde\Lambda(1)=1$ and $\tilde\Lambda(\tau(\mu g))=0$ for all generators $g$ and monomials $\mu$ with $\deg(\mu g)\le D$, then $D\le N+d_G-1$. Hence a pullback through $\tau$ proves at most $\degns(E)\ge N+d_G$.
\end{theorem}
\begin{IEEEproof}
Let $G$ be the generators that are not Boolean axioms. By~(C1) fix Boolean variables $v_k$ with $\tau(v_k)=\zeta_k$. For a monomial $\zeta^\alpha$, let $v^\alpha$ be the corresponding monomial in the $v_k$; it has the same degree, and $\tau(v^\alpha)=\zeta^\alpha$.

\emph{Boolean multiples.} Since $\Bool{\zeta_k}=\tau(\Bool{v_k})$, we get $\tilde\Lambda(\zeta^\alpha\Bool{\zeta_k})=\tilde\Lambda(\tau(v^\alpha\Bool{v_k}))=0$ whenever $|\alpha|+2\le D$. So $\tilde\Lambda(a\Bool{\zeta_k})=0$ whenever $\deg(a\Bool{\zeta_k})\le D$.

\emph{No Boolean zero of the image.} Suppose $p\in\{0,1\}^N$ satisfied $\tau(g)(p)=0$ for all $g\in G$. Let $v^*$ be the point with coordinates $\tau(v)(p)$. Then $g(v^*)=\tau(g)(p)=0$ for $g\in G$, and $\Bool v(v^*)=\tau(\Bool v)(p)=0$ by~(C2). So $v^*$ would be a common zero of $E$.

\emph{A certificate for the image.} Hence $F=\sum_{g\in G}\tau(g)^2$ is positive on $\{0,1\}^N$. Let $h_g$ be the multilinear polynomial equal to $\tau(g)/F$ on the cube, so $\deg h_g\le N$. As in the proof of Proposition~\ref{prop:generic-upper}, $1-\sum_gh_g\tau(g)=\sum_ka_k\Bool{\zeta_k}$ with $\deg(a_k\Bool{\zeta_k})\le N+d_G$, by~(C3).

\emph{Conclusion.} Suppose $D\ge N+d_G$. For every monomial $\zeta^\alpha$ of $h_g$, $\tilde\Lambda(\zeta^\alpha\tau(g))=\tilde\Lambda(\tau(v^\alpha g))=0$, because $\deg(v^\alpha g)\le N+d_G\le D$. Also $\tilde\Lambda(a_k\Bool{\zeta_k})=0$ by the first step. Applying $\tilde\Lambda$ to $1=\sum_gh_g\tau(g)+\sum_ka_k\Bool{\zeta_k}$ gives $1=0$. Hence $D\le N+d_G-1$.
\end{IEEEproof}

\begin{corollary}[Optimality of the planting method]\label{cor:cap-planted}
All generators of $E^{\cnt}_{t,b}(H)$ other than Boolean axioms are linear. So a pullback through a substitution satisfying (C1)--(C3) into $N$ Boolean variables proves at most $\degns(E^{\cnt}_{t,b}(H))\ge N+1$. The constant-selector substitution of Theorem~\ref{thm:B} satisfies (C1)--(C3) with $N=M$: each $\eta_v$ is the image of a support variable, every Boolean variable maps to $0$, $1$, $\eta_v$ or $1-\eta_v$, so every Boolean axiom maps to $0$ or to some $\Bool{\eta_v}$, and every other generator maps to a polynomial of degree at most one. Theorem~\ref{thm:B} and all its corollaries are therefore optimal among such pullbacks. In particular:
\begin{enumerate}
\item[(a)] adding further polynomial constraints on the same $M$ knapsack variables cannot raise the target degree above $\degns(\Tgt_{M,r})=M+1$, since enlarging a system can only lower its degree;
\item[(b)] circulant or quasi-cyclic structure does not enter the bound, which depends only on $N$;
\item[(c)] if (C3) is weakened to $\deg\tau(g)\le\delta d_G$, for a substitution of degree $\delta$, the same proof gives $D\le N+\delta d_G-1$.
\end{enumerate}
\end{corollary}

The selector-carried substitution of Remark~\ref{rem:selector} does not satisfy (C2), since the Boolean axiom of $y_c$ maps to $T$ times a linear polynomial; as it covers the same instances, Corollary~\ref{cor:cap-planted} applies to Theorem~\ref{thm:B} as stated. Independently, the target system $\Tgt_{M,r}$ has degree exactly $M+1$ (Proposition~\ref{prop:target}).

\subsection{Sharpness of the Planted Bound}\label{sec:sharp}
\begin{proposition}[A sharp family]\label{prop:sharp}
For $M\ge1$ let $F_M$ have $2M+2$ checks and $M+2$ columns: columns $o_1$ and $o_2$ of weight one in distinct checks, and columns $p_0,\dots,p_{M-1}$ of weight two, every check having degree one. Take $t=1$, $b=0$, $O=\{o_1,o_2\}$ and planted columns $p_0,\dots,p_{M-1}$. Then (P1)--(P5) hold with $w_*=2$ and $\beta=0$, problem~\eqref{eq:problem} is infeasible, and
\[
\degns\big(E^{\cnt}_{1,0}(F_M)\big)=M+1 .
\]
\end{proposition}
\begin{IEEEproof}
(P1) and (P2) are immediate. (P3) and (P4) hold because every check meets exactly one column, and $\Phi(O)=2=w_*$ gives (P5). Every singleton has $\Phi\in\{1,2\}$, so the problem is infeasible, and Corollary~\ref{cor:planted} gives $\degns\ge M+1$.

For the upper bound, every check $j$ has degree one, so $K_j=0$ and $L_j=x_{\iota(j)}-u_j$, where $\iota(j)$ is the column of $j$. Put $Y'=\sum_vx_{p_v}$; then $\sum_ix_i=x_{o_1}+x_{o_2}+Y'$ and $\sum_jL_j=x_{o_1}+x_{o_2}+2Y'-\sum_ju_j$. For $t=1$ and $b=0$ the generators $W-\sum_\ell\ell y_\ell$ and $U-\sum_\varrho\varrho z_\varrho$ of~\eqref{eq:cnt} are $W-y_1$ and $U$, so
\begin{align*}
\textstyle\sum_ju_j&=U-(U-\textstyle\sum_ju_j),\\
\textstyle\sum_ix_i-1&=(W-y_1)-(W-\textstyle\sum_ix_i)+(y_0+y_1-1)-y_0,
\end{align*}
and therefore
\[
Y'+1=\textstyle\sum_jL_j+\sum_ju_j-\big(\sum_ix_i-1\big)
\]
is a combination of generators of degree one with constant coefficients. Proposition~\ref{prop:target}(a) with $r=-1$, applied to the variables $x_{p_v}$, gives $1=f_{M,-1}\cdot(Y'+1)+\sum_vh_v\Bool{x_{p_v}}$ with products of degree at most $M+1$. Substituting the expression for $Y'+1$ gives a certificate of degree $M+1$.
\end{IEEEproof}

\emph{Where equality occurs.} In $F_M$ every planted column lies in checks of degree one, so its parity variables are linear in the support variables. Exact computations settle the small cases. We took all matrices with at most three checks and at most six columns, or four checks and at most five columns, with no zero column, up to row and column permutations; zero rows are allowed (they do not change the degree, since $u_j=-L_j$ for a zero row). With $t\le2$ and $b\le2$ this gives $423$ infeasible instances that admit a planting satisfying (P1)--(P5). All have $b=0$ ($301$ at $(t,b)=(1,0)$ and $122$ at $(2,0)$), and the largest planting has $M=1$ ($402$ instances) or $M=2$ ($21$ instances). The exact degree of every instance over $\Q$ (Section~\ref{sec:verif}) equals $M+1$ in $4$ instances (one with $M=1$, three with $M=2$), $M+2$ in $97$ ($79$ with $M=1$, $18$ with $M=2$), and $M+3$ in $293$ and $M+4$ in $29$ instances, all with $M=1$; so the degrees range from $2$ to $5$, and $4+97+293+29=423$. In each of the four instances with equality, every check of every planted column has degree one; the converse fails in $69$ further instances, in which every maximum planting has this property and the degree exceeds $M+1$. The next subsection shows that equality does not require checks of degree one once $M\ge3$.

\subsection{Partners}
Say that a planting satisfies the \emph{partner condition} if every check of every planted column contains a second column, necessarily outside $O$ and the planted set by (P3) and (P4). Let $(t,b)=(1,0)$ and $M\ge1$, and let $H^\star_M$ have checks $a_1,a_2,c_0,\dots,c_{M-1}$ and columns $o_1,o_2,p_0,\dots,p_{M-1},g$ with
\[
N(a_1)=\{o_1,o_2\},\quad N(a_2)=\{o_2\},\quad N(c_v)=\{p_v,g\},
\]
where $v\in\ZM$.

\begin{proposition}[Partners do not force degree $M+2$]\label{prop:partner-cex}
For $H^\star_M$ with $O=\{o_1,o_2\}$ and planted columns $p_0,\dots,p_{M-1}$, hypotheses (P1)--(P5) hold with $w_*=1$ and $\beta=0$, the partner condition holds with the partner $g$ in every planted check, and problem~\eqref{eq:problem} is infeasible. Moreover,
\[
M+1\le \degns\bigl(E^{\cnt}_{1,0}(H^\star_M)\bigr)\le\max\{M+1,4\}.
\]
In particular, $\degns=M+1$ for every $M\ge3$: the bound of Corollary~\ref{cor:planted} is attained although every planted check has degree two.
\end{proposition}
The proof, which shows that the partner and $O$ are pinned at degree four and that the knapsack on the planted columns then proceeds as in $F_M$, is given in Section~S-II of the supplementary material. Exact computation gives $\degns(E^{\cnt}_{1,0}(H^\star_M))=3,4,4$ for $M=1,2,3$, so the overhead of the partner is visible only for $M\le2$.

In $H^\star_M$ the partner $g$ has weight $M$, so Proposition~\ref{prop:partner-cex} does not address codes of bounded degree; but bounded degree alone does not rule out degree $M+1$ either. Let $G_M$ be obtained from $H^\star_M$ by giving each planted check $c_v$ a private partner $k_v$ with $C(k_v)=C(p_v)=\{c_v\}$, instead of $g$. Then all column weights and check degrees are at most two, and the Tanner graph is acyclic. Exact computations, reduced by the symmetry group of order $2^MM!$ that permutes the blocks $\{p_v,k_v,c_v\}$ and swaps $p_v$ with $k_v$ (Section~\ref{sec:verif}), give
\[
\degns\big(E^{\cnt}_{1,0}(G_M)\big)=\begin{cases}M+2,&1\le M\le3,\\ M+1,&4\le M\le7.\end{cases}
\]
These structures are degenerate: they use weight-one columns, a degree-one check and repeated columns. The following question excludes all three.

\begin{problem}[Partners in nondegenerate bounded-degree LDPC structure]\label{prob:bounded-partner}
Assume (P1)--(P5) and the partner condition. Suppose there are constants $w_{\max}$ and $d_{\max}$, independent of $M$, such that $2\le|C(i)|\le w_{\max}$ for every column $i$ and $2\le d_j\le d_{\max}$ for every check $j$. Assume in addition that no two columns have identical check neighbourhoods. Can $\degns(E^{\cnt}_{t,b}(H))=M+1$ occur for arbitrarily large $M$? In particular, what changes for column-regular matrices, or under Tanner girth at least six?
\end{problem}

Small partner-planted instances never attain $M+1$. Among all matrices with $m=3$ checks and $n\le7$ columns, $m=4$ and $n\le6$, or $m=5$ and $n\le4$, without zero rows or columns and up to permutations, and all $(t,b)$ with $t\le3$ and $b\le2$, exactly $1409$ instances are infeasible and admit a planting satisfying (P1)--(P5) and the partner condition; with $M^*$ the largest such planting ($M^*=1$ in $1324$ instances and $M^*=2$ in $85$), an exact rational functional shows $\degns\ge M^*+2$ in all of them. A counterexample with $M=3$ has $m\ge5$ and $n\ge6$: it needs three disjoint planted checks, at least two checks in $C(O)$ (a single check would force identical or zero columns in $O$), three planted columns, two columns of $O$ and a partner; so $H^\star_3$ is minimal among them.

\subsection{Cycle Codes}\label{sec:sharp-cycle}
On $C_n$ the data show a gap of exactly $2$. Exactly, $\degns(E^{\cnt}_{1,0}(C_n))=4,4,5,5$ for $n=5,\dots,8$ (Table~\ref{tab:mixed}), in each case $2$ above $M+1=\lfloor(n-1)/2\rfloor$; and $\degns(\OPP_n)=\lceil n/2\rceil+1$ exceeds the bound of Corollary~\ref{cor:lin-opp} by exactly $2$ for $5\le n\le12$. Single-column plantings are maximal at $M=\lfloor(n-3)/2\rfloor$ (Corollary~\ref{cor:lin-cycle}(b)), so by Corollary~\ref{cor:cap-planted} no such pullback gives more than $\lfloor(n-1)/2\rfloor$. Reaching the conjectured value $\lceil n/2\rceil+1$ by a pullback needs at least $\lceil n/2\rceil$ target variables, for instance checks with two planted literals whose parities are carried by additional target variables. We have not found such a target of maximal degree.

\subsection{Further Limitations}
(i)~\emph{Infeasibility} is not implied by Theorem~\ref{thm:B} and must be established separately; for $b=0$ and $t\ge\dmin$ the problem is feasible, and for $t\ge4$ in Theorem~\ref{thm:triangle} it depends on more than the girth.
(ii)~\emph{Characteristic zero} is essential (Remark~\ref{rem:planted-hyp}).
(iii)~\emph{Other plantings.} Weighted knapsacks and plantings with two planted columns in one check, which cost the factor $\frac12$ of Theorem~\ref{thm:red}, are not explored.
(iv)~\emph{Mixed encodings.} The planted substitution sends some local obstruction to a nonzero polynomial (Section~\ref{sec:mixed}), so it gives nothing for the mixed encodings.
(v)~\emph{Odd column weight} at $(t,b)=(1,0)$ remains open for girth-six families without induced paths, stars or many check-disjoint short codewords, such as the projective-plane codes of Proposition~\ref{prop:exist}(c).

\section{Exact Computational Verification}\label{sec:verif}
The proofs above are complete; the computations confirm them on instances, supply the exact values in Table~\ref{tab:mixed}, in the examples and in the census of Section~\ref{sec:sharp}, and produce the certificates of Table~\ref{tab:qc}. The scripts of the supplementary material reproduce them; the supplementary package also contains the matrices, certificates and configurations and a README with requirements and run times, and Section~S-III of the supplementary document gives the full list of computations and, in Table~S1, the script for each claim. Every Nullstellensatz degree or degree bound obtained by computation rests on a certificate or functional verified exactly over $\Q$; minimum distances are found by exhaustive enumeration, configurations are verified exactly, and packing optima are proved by optimization solvers.

\begin{table*}[t]
\caption{Local-encoding certificates for QC-LDPC matrices, built with the construction of Theorem~\ref{thm:A}}\label{tab:qc}
\centering\scriptsize\setlength{\tabcolsep}{4.0pt}
\begin{tabular}{lcccrrccccc}
\hline
Matrix & $n$ & $(t,b)$ & $K$ & $|E^{\loc}|$ & $|E^{\cnt}|$ & bound $t{+}b{+}K{+}1$ & constructed deg. & terms & time (s) & file size\\ \hline
Regular, $Z=8$   & 48 & $(3,0)$ & 2 & 194{,}749 & 156 & 6 & 6 & 639{,}853 & 14.6\,/\,6.7\,/\,2.4 & 19.3~MB\\
Irregular, $Z=8$ & 48 & $(3,0)$ & 2 & 194{,}749 & 156 & 6 & 6 & 556{,}914 & 12.8\,/\,5.6\,/\,2.0 & 17.1~MB\\
Irregular, $Z=8$ & 48 & $(2,1)$ & 2 & 17{,}717  & 156 & 6 & 6 & 153{,}038 & 0.9\,/\,1.1\,/\,0.5 & 4.5~MB\\
Regular, $Z=16$  & 96 & $(2,2)$ & 2 & 160{,}465 & 301 & 7 & 7 & 1{,}525{,}118 & 8.1\,/\,16.4\,/\,6.3 & 53.8~MB\\
5G NR BG2, $Z=2^\dagger$ & 104 & $(2,0)$ & 3 & 182{,}625 & 447 & 6 & 6 & 162{,}398 & 2.3\,/\,1.2\,/\,0.4 & 4.5~MB\\ \hline
\end{tabular}\\[2pt]
\parbox{\textwidth}{\scriptsize $|E^{\loc}|$ and $|E^{\cnt}|$: generator counts~\eqref{eq:countloc} and~\eqref{eq:countcnt}. Constructed deg.: degree of the certificate the construction produces, an upper bound on the Nullstellensatz degree, not its exact value. Terms: monomials in the stored multipliers, after like terms are collected within each multiplier. Time: construction / Python verifier / C++ verifier, in seconds, wall clock on one core of a shared Intel Xeon at 2.3~GHz (Python~3.13; g++~13.3 with \texttt{-O2}); the construction time excludes writing the file, and the verification times include reading it. File size: uncompressed certificate file. $^\dagger$Stress test at the feasibility boundary, not a practical code. The certificates of~\cite{itw27}, built with the former bound, had constructed degree $7,7,7,8$ and $1{,}221{,}213$, $1{,}053{,}733$, $240{,}806$ and $1{,}973{,}966$ terms.}
\end{table*}

\subsection{Exact Degree Computations}
For systems whose variables are all Boolean, Lemma~\ref{lem:mlred} reduces the question $1\in I_D$ to linear algebra over $\Q$ on the rows $\ML(\mu g)$. For counting encodings we use $E'$ of Lemma~\ref{lem:elim-gen}, and, when at most one selector remains on each side, the following smaller system.

\begin{lemma}[Selector elimination]\label{lem:elim}
Let $t\le2$ and $b\le1$, and let $E^*$ be obtained from $E'$ of Lemma~\ref{lem:elim-gen} by the substitution $y_0\mapsto0$, $y_1\mapsto1-\sum_{2\le\ell\le t}y_\ell$, $z_0\mapsto1-\sum_{1\le\varrho\le b}z_\varrho$. Then $E^*$ consists of the $L_j$, the generators $1+\sum_{2\le\ell\le t}(\ell-1)y_\ell-\sum_ix_i$ and $\sum_{1\le\varrho\le b}\varrho z_\varrho-\sum_ju_j$, and the Boolean axioms of the remaining variables, and $\degns(E^*)=\degns(E^{\cnt}_{t,b}(H))$. The same substitution, applied to $\Emix{XY}{1,0}(H)$ after eliminating its counters, removes all selectors and leaves $\sum_ix_i-1$ if the support is counted and $\sum_ju_j$ if the syndrome is counted.
\end{lemma}
\begin{IEEEproof}
The substitution is affine and does not raise degrees. Since at most one selector remains on each side, it maps $\Bool{y_1}$ to $\Bool{y_2}$ or $0$, $\Bool{z_0}$ to $\Bool{z_1}$ or $0$, $\Bool{y_0}$, $y_0$, $\sum_\ell y_\ell-1$ and $\sum_\varrho z_\varrho-1$ to $0$, and the two counting generators to the generators of $E^*$; so certificates for $E'$ map to certificates for $E^*$ of no larger degree. Conversely, $1+\sum_{\ell\ge2}(\ell-1)y_\ell-\sum_ix_i=(\sum_\ell\ell y_\ell-\sum_ix_i)-(\sum_\ell y_\ell-1)+y_0$ and $\sum_\varrho\varrho z_\varrho-\sum_ju_j$ are constant combinations of degree-one generators of $E'$. Lemma~\ref{lem:elim-gen} completes the proof; for the mixed encodings the argument is the same.
\end{IEEEproof}

Each degree test at a given $D$ computes a candidate certificate (rational multipliers $c_k$ with $\sum_kc_k\ML(\mu_kg_k)=1$) or a candidate functional as in Lemma~\ref{lem:dual} modulo word-size primes, reconstructs it over $\Q$ by Chinese remaindering and rational reconstruction, and then \emph{verifies it exactly}: a certificate by expanding the sum over $\Q$, a functional by checking $\Lambda(1)=1$ and $\Lambda(\ML(\mu g))=0$ on every row. The modular computation is only a search heuristic. A rank modulo $p$ can be smaller than the rank over $\Q$, so a membership test modulo a single prime can err in either direction; no result in this paper rests on one. All moduli are certified primes. For $C_n$ the dihedral group, and for the family $G_M$ of Section~\ref{sec:limits} the group $\mathbb Z_2\wr S_M$ of order $2^MM!$, permute the generators; a certificate or functional of degree at most $D$ exists if and only if an invariant one does, so we compute invariant candidates on monomial orbits and verify them exactly in the same way (for $G_M$ the reduced computation agrees with the unreduced one for $M\le3$).

These methods give the exact values of Propositions~\ref{prop:tight} and~\ref{prop:B} (local degree $2$ for $C_3,\dots,C_8$, $\mathsf K_4$, the Fano plane and the Petersen graph), of $\degns(\OPP_n)$ for $4\le n\le12$, of every entry of Table~\ref{tab:mixed}, of the $423$ census instances, of $F_M$ and $H^\star_M$ for $M\le3$ and of $G_M$ for $M\le7$, and the partner census of Section~\ref{sec:limits}. The census first records, for each instance, an exact functional at $D=M$, confirming $\degns>M$ independently of Corollary~\ref{cor:planted}, and then increases $D$ until an exact certificate is found.

\subsection{Identity and Planting Checks}
Every polynomial identity in Sections~\ref{sec:local}, \ref{sec:planting} and~\ref{sec:mixed} was checked in exact arithmetic over the ranges listed in Section~S-III. The construction of Theorem~\ref{thm:A} was executed literally on $165$ random infeasible instances, $40$ of them with $b=2$: every certificate satisfies $\sum_ga_gg=1$ exactly, has degree at most $t+b+K+1$, attained in $161$ cases, and has exactly the number of multiplier terms counted in the proof, at most $0.361$ times the bound~\eqref{eq:size}. For Theorem~\ref{thm:B}, every generator of the full counting encoding was mapped through $\tau$ in $1500$ random instances of (R1)--(R4), with $|P_v|\le3$ and all $(t,b)\in\{1,2,3\}\times\{0,1,2\}$, and in $1450$ instances with $r\in\{0,\dots,M\}$ problem~\eqref{eq:problem} is feasible, as shown after Theorem~\ref{thm:B}. On the full counting encodings of nine planted configurations, counters and selectors included, $\Lambda_{M,r}\circ\tau$ vanishes on every product $\mu g$ of degree at most $M$ (up to $567{,}604$ products) and not on every product of degree $M+1$; the planted-codeword and domino cases were run with both substitutions of Remark~\ref{rem:selector}. For the examples of Sections~\ref{sec:md} and~\ref{sec:odd}, $\dmin$ was found by exhaustive search for codewords through the first column of each column block, which covers all codewords up to cyclic shifts; the maximum packings were computed with the CP-SAT solver of OR-Tools and reported as optimal only when the solver proved optimality; and every bound was checked generator by generator. A separate implementation, sharing no code with these scripts, recomputes $\dmin$, the minimum-weight codewords and the configurations, and recomputes every packing optimum as an integer program solved by HiGHS; all values agree.

\subsection{QC-LDPC Certificates}
We generated local-encoding certificates for three quasi-cyclic matrices lifted from $3\times6$ base matrices: a $(3,6)$-regular one lifted with $Z=8$ and $Z=16$, and an irregular one with two zero blocks, column weights $2$ and $3$ and check degrees $5$ and $6$, lifted with $Z=8$. All three have girth~$6$ and $K=2$; their exponent matrices are listed in Section~S-III. As a stress test of the pipeline at the feasibility boundary we also lifted base graph~2 of 5G NR~\cite{3gpp38212}, which has $42$ checks, $52$ columns and $38$ degree-one columns, with the smallest lifting size $Z=2$, from lifting set~$0$, so that the exponents are the set-$0$ entries of Table~5.3.2-3 of~\cite{3gpp38212} reduced modulo~$2$. We read these entries from two independent machine-readable transcriptions of the table, in Sionna~0.19.2~\cite{sionna} and in release~23.11 of srsRAN~4G~\cite{srsran4g}, which agree on all $197$ entries in all eight lifting sets. The lifted matrix has $n=104$, $m=84$, $76$ columns of weight one and check degrees from $3$ to $10$, so $K=3$. It is not a code one would use: a weight-one column alone has $\Phi=1$, so problem~\eqref{eq:problem} is feasible for every $b\ge1$. We take $(t,b)=(2,0)$, which is infeasible while $(2,1)$ is feasible. The example tests the certificate generator and the verifiers at the feasibility boundary; it says nothing about the performance of 5G NR codes. Exhaustive enumeration confirms that problem~\eqref{eq:problem} is infeasible for every case in Table~\ref{tab:qc}, and the generator counts agree with~\eqref{eq:countcnt} and~\eqref{eq:countloc}.

The generator writes every multiplier directly from~\eqref{eq:key}, \eqref{eq:A2} and Lemma~\ref{lem:absorb}, with no Gr\"obner basis and no reduction of a residual. Each certificate attains the degree bound $t+b+K+1$. For the four cases of~\cite{itw27} this is one less than the degree of the certificates there, and the new certificates have between $23\%$ and $48\%$ fewer terms. Before like terms are collected, the multipliers of the five certificates contain $4.41$, $3.93$, $0.32$, $2.36$ and $0.82$ million monomials, that is, $0.44$, $0.39$, $0.49$, $0.62$ and $0.18$ times the bound~\eqref{eq:size}. Regenerating the five certificates reproduces the published files byte for byte. Each certificate stores $H$, $(t,b)$, a SHA-256 hash of $H$ and the rational multipliers, and is checked by two verifiers that share no code with each other or with the generator, one in Python with exact rational arithmetic and one in C++ with GMP. Each verifier rebuilds the generators from $H$, $t$ and $b$, checks the hash and every descriptor, and expands $\sum_ga_gg-1$ exactly. Both accept all five certificates, and both reject every one of $48$ corrupted files, $16$ for each of three certificates.

\section{Discussion and Open Problems}\label{sec:disc}
\emph{Size versus degree.} The local encoding trades generators for degree: $\binom{n}{t+1}+\binom{m}{b+1}=O(n^{t+1}+m^{b+1})$ obstructions and $O(n^t)$ certificate terms (Theorem~\ref{thm:A}) buy a degree independent of $n$ for fixed $t$, $b$ and bounded check degree, whereas the counting encoding has $O(n+m+\sum_jK_j)$ generators and, by Sections~\ref{sec:sep}--\ref{sec:odd}, degree $\Theta(n)$ on cycle codes, on regular girth-six families near the feasibility boundary, on array codes, and for certifying the minimum distance of arrays of circulants. Neither encoding is cheaper in every respect, and proof degree is not a running time.

\emph{Encoding sensitivity as a design rule.} The support obstructions $x_T$ are redundant at the level of ideals, but they enable the explicit low-degree certificate construction of Theorem~\ref{thm:A}, in the spirit of the degree cutters of~\cite{deloera11}; whether equally low-degree certificates exist without them remains open. On cycle codes at $(t,b)=(1,0)$, Section~\ref{sec:mixed} shows that a single global count is harmless, while two counts allow a knapsack to be planted. This suggests writing cardinality constraints locally when certificates are a deliverable.

\emph{Open problems.}
\begin{enumerate}
\item The exact values of $\degns(\OPP_n)$ (Conjecture~\ref{conj}) and of $\degns(E^{\cnt}_{1,0}(C_n))$; Table~\ref{tab:mixed} suggests that they coincide, so the factor $\tfrac12$ in Theorem~\ref{thm:red} is probably an artifact.
\item Whether the local encoding without the support obstructions $x_T$ still admits certificates of degree independent of the block length (Section~\ref{sec:local}).
\item Whether $\degns(E^{\loc}_{t,b}(H))\ge t+b+1$ whenever all check degrees are at least two, and whether degree two persists for column-regular matrices with check degree at least four (Section~\ref{sec:local}).
\item Whether $\degns(\Emix{CL}{1,0}(C_n))=4$ for every $n\ge6$, and the behaviour of the mixed encodings beyond $C_n$ and $(t,b)=(1,0)$.
\item Whether $\degns(E^{\cnt}_{1,0}(H))$ grows for column-regular families of odd column weight that contain no induced paths, stars or linearly many check-disjoint short codewords, for example the projective-plane codes (Proposition~\ref{prop:exist}(c)).
\item The partner question of Open Problem~\ref{prob:bounded-partner}.
\item How far the exact degree of the counting encoding can exceed the planted bound $M+1$, for example in the minimum-distance certificates of Examples~\ref{ex:A} and~\ref{ex:B}; by Corollary~\ref{cor:cap-planted}, a larger lower bound needs a different kind of argument.
\end{enumerate}

\section{Conclusion}\label{sec:concl}
Two exact polynomial encodings of the same syndrome-exclusion problem behave very differently. The local encoding has explicit Nullstellensatz certificates of degree at most $t+b+K+1$, independent of the block length for fixed $t$, $b$ and bounded check degree. For the counting encoding, the generalized planted-knapsack theorem gives lower bounds linear in the block length on cycle, regular and array-based QC-LDPC codes, for certifying the minimum distance of arrays of circulant permutation matrices, and, through planted pairs on induced paths or stars, for odd column weight with no unsatisfied checks. The planting argument is optimal among pullbacks of its kind and its bound is attained; the exact degree of the counting encoding beyond the planted bound remains open.

\end{document}